\RequirePackage{etoolbox}
\csdef{input@path}{%
 {sty/}
 {img/}
}%
\csgdef{bibdir}{bib/}

\documentclass[12pt]{article}
\usepackage{xr}
\usepackage[margin=0.5in]{geometry}
\usepackage{amsthm}
\usepackage{amsmath}
\usepackage{natbib}
\usepackage[colorlinks,citecolor=blue,urlcolor=blue,filecolor=blue,backref=page]{hyperref}
\usepackage{graphicx}
\usepackage{changepage}
\usepackage[dvipsnames]{xcolor}

\usepackage{mathrsfs}
\usepackage{appendix}
\usepackage{subcaption}
\usepackage{float}
\usepackage{caption}
\usepackage{multirow}
\bibpunct{(}{)}{;}{a}{}{,}

\newtheorem{theorem}{Theorem}

\newcommand{\given}{\,|\,}
\newcommand{\eps}{\epsilon}

\newcommand{\calK}{{\cal K}}

\newcommand{\calD}{{\cal D}}
\newcommand{\calG}{{\cal G}}
\newcommand{\calT}{{\cal T}}

\newcommand{\calO}{{\cal O}}

\newcommand{\ind}{\stackrel{\mathrm{ind}}{\sim}}

\newcommand{\bitem}{\begin{itemize}}
	\newcommand{\eitem}{\end{itemize}}

\title{Spatial disease mapping using directed acyclic graph auto-regressive (DAGAR) models}
\author{Abhirup Datta\\ Johns Hopkins University
	\and Sudipto Banerjee\\ University of California Los Angeles
\and James S. Hodges\\ University of Minnesota
\and Leiwen Gao\\ University of California Los Angeles}
\begin{document}
\maketitle
%
%
%
%
%
%



\begin{abstract}
		Hierarchical models for regionally aggregated disease incidence data commonly involve region specific latent random effects that are modeled jointly as having a multivariate Gaussian distribution. The covariance or precision matrix incorporates the spatial dependence between the regions. Common choices for the precision matrix include the widely used ICAR model, which is singular, and its nonsingular extension which lacks interpretability. We propose a new parametric model for the precision matrix based on a directed acyclic graph (DAG) representation of the spatial dependence. Our model guarantees positive definiteness and, hence, in addition to being a valid prior for regional spatially correlated random effects, can also directly model the outcome from dependent data like images and networks. Theoretical results establish a link between the parameters in our model and the variance and covariances of the random effects. Substantive simulation studies demonstrate that the improved interpretability of our model reaps benefits in terms of accurately recovering the latent spatial random effects as well as for inference on the spatial covariance parameters. Under modest spatial correlation, our model far outperforms the CAR models, while the performances are similar when the spatial correlation is strong. We also assess sensitivity to the choice of the ordering in the DAG construction using theoretical and empirical results which testify to the robustness of our model. We also present a large-scale public health application demonstrating the competitive performance of the model. 

\vskip 5mm 
\noindent \textbf{Keywords:} Areal data, Bayesian inference, Directed acyclic graphs, Disease mapping, Spatial autoregression
\end{abstract}





	\section{Introduction}\label{sec:intro}
\noindent Epidemiological data for disease rates are often presented as aggregated disease counts over entire geographical regions like states or counties. Such \emph{areal} or \emph{areally-referenced} data are  ubiquitous in public health applications. Accurate identification of trends and factors associated with the disease requires accounting for the spatial dependence among the regions. A common approach to analyze areal datasets envisions the geographic domain as an undirected graph with the regions constituting the vertices and an edge between two vertices if the corresponding regions share a geographical border. This creates well defined neighbors for each region which are used to specify the joint or conditional distributions of region-specific latent Gaussian random effects in a hierarchical setup. For example, the popular conditional autoregressive (CAR) model \citep{besag74,clay92} incorporates the underlying neighborhood structure in specifying the full conditional distribution for each observation.  
If $w_i$ denotes the random effect representing the $i$th region for $i=1,\ldots,k$  and $i \sim j$ indicates that regions $i$ and $j$ are neighbors, then the CAR model specifies the full conditional distributions
\begin{equation}\label{eq: car}
w_i \mid w_{-i} \sim N\left(\sum_{j \sim i} w_j/n_i, \tau_w n_i\right)\;,
\end{equation}
where $w_{-i}$ denotes the vector of observations leaving out the $i$th one, $n_i$ denotes the number of neighbors for the $i$th region and throughout the text we adopt the convention that $N(\alpha,\Delta)$ denotes normal distribution with mean $\alpha$ and precision $\Delta$, both in univariate and multivariate contexts. Hence, in (\ref{eq: car})  above, $\tau_w n_i$ is the conditional precision of $w_i \mid w_{-i}$.

The joint distribution of $w= (w_1,\ldots,w_k)^T$ can be derived from (\ref{eq: car}) as $w \sim N(0, \tau_w (D-A))$ where $A=(a_{ij})$ is the adjacency matrix of the neighborhood graph i.e. $a_{ij} =1$ if and only if $i \sim j$, and $D$ is a diagonal matrix with $n_1,\ldots,n_k$ on the diagonal. 
As $D-A$ is singular, this construction yields an improper joint distribution of the $w_i$'s, referred to as the intrinsic or improper CAR (ICAR) model. 
	This impropriety renders the model ineligible for directly modeling the response or for generating data, although both can proceed by using contrasts as demonstrated in \citet{kooper}. Also, the distribution can still be used as a prior for latent spatial random effects $w$ and the posterior of $w$ usually remains valid. 

The impropriety of the ICAR model can be rectified by generalizing the full conditional mean to $E(w_i \mid w_{-i}) = \rho \sum_{j \sim i} w_j /n_i$ yielding the joint distribution $w \sim N(0, \tau_w (D-\rho A))$ which is proper for a certain range of $\rho$. Although introduction of $\rho$ imparts more flexibility than the parameter-free improper analogue, it is difficult to interpret $\rho$ as even very high values of $\rho$ induce only modest spatial correlation among the observations \citep[see][for a discussion on this]{ban14}. Furthermore, \cite{wall04} shows that even negative values of $\rho$ may lead to positive correlation among neighboring regions. \cite{assun09} 
found that these oddities are a general feature of CAR models.

The second popular approach is the simultaneous autoregressive (SAR) model \citep{whittle54} which proceeds by simultaneously modeling the random effects as
\begin{equation}\label{eq:sar}
w_i = \sum_{j \neq i} b_{ij} w_j + \eps_i \mbox{ for } i=1,2,\ldots,k
\end{equation}
where $\eps_i \ind N(0,\tau_i)$ are errors independent of $w$. Defining $B=(b_{ij})$ and $F$ to be a diagonal matrix with entries $\tau_1,\ldots,\tau_k$, the set of equations in (\ref{eq:sar}) 
yields the joint distribution $w \sim N(0,(I-B)F(I-B)^{T})$. However, the common choice of defining $b_{ij}=\rho I(i \sim j) / n_i$, where $I(\cdot)$ denotes the indicator function,  
leads to similar problems with respect to interpretation of the parameter $\rho$ \citep{wall04}. 

Beyond these two approaches, the inventory of covariance models for areal datasets is very limited. \cite{leroux00} and \cite{mac00} extended the CAR model by accommodating over-dispersion alongside spatial information. They proposed using the precision matrix $\lambda \tau_w (D-A) + (1-\lambda) \tau_w I$, where $\lambda \in [0,1]$ controls the degree of dependence among the regions. For a regular graph where all vertices have same number of neighbors $d$, $D=dI$. In this case, $\lambda \tau_w (D-A) + (1-\lambda) \tau_w I$ can be rewritten as $ \frac {1+ (d-1)\lambda}d \tau_w (D-\rho^* A )$ where $\rho^* = \frac {d\lambda}{1+ (d-1)\lambda}$. Thus, if the numbers of neighbors for the vertices do not vary greatly, this approach is somewhat similar to the proper CAR model and is encumbered by the same interpretability concerns. 

	For lattice based applications, there is a richer class of parametric intrinsic autoregression models \citep{kooper,higdon99}. 
	However, all such intrinsic models rely heavily on the lattice structure and cannot be used directly for arbitrary graphs. Applications to irregular areal data can proceed by breaking up the region into a fine lattice, using the intrinsic model on the lattice and aggregating over each area. \citet{mondal05} demonstrated that certain classes of intrinsic autoregressive models can be interpreted as 
	average of a fine scale Gaussian Process over the entire domain. 
	In disease mapping contexts, where the data are often observed over fixed politically delineated regions,
    such a latent fine scale spatial process may be difficult to interpret. In this manuscript, we only focus on models that can be formulated directly on the areal units.

We propose a new way of constructing precision matrices for areal models using a directed acyclic graph derived from the original undirected graph. Directed acyclic graphs or {\em DAGs} have been used in the spatial literature for modeling large spatial datasets \citep{nngp} and for generating image textures \citep{pomm}. Instead of modeling the precision matrix directly, 
we model its Cholesky factor, which for any multivariate Gaussian distribution is determined by the conditional distributions of the $w_i$'s. We specify these conditional distributions using autoregressive covariance models on a sequence of local trees created from this directed acyclic graph. The resulting Cholesky factor and the precision matrix are sparse. We refer to this model as the directed acyclic graph autoregressive or DAGAR model.  

Unlike the ICAR model, our model's covariance matrix is guaranteed to be positive definite. This opens up a new avenue to generate or directly model multivariate Gaussian data with dependence structure derived from a graph. Common examples of such data, besides aggregated regional data, include images or social network data. We establish, both theoretically and empirically, that our model endows $\rho$ with a clear interpretation as a spatial autocorrelation parameter, which, in fact, resolves an important conundrum in the conditional and simultaneous autoregressive models \citep[][]{wall04}. Also, the Cholesky factor has the same level of sparsity as the undirected graph ensuring scalability for analyzing very large areal datasets. 

Cholesky factors inherit the dependence of directed acyclic graphs on ordering of the regions, thereby making our model order-dependent. As spatial regions generally do not have any natural ordering, to understand the impact of ordering, we propose a novel order free model  by averaging over all $k!$ possible orderings. We show that the resulting precision matrix, which is order-free, can be evaluated in closed form and we use it to present some theoretical results suggesting that the DAGAR precision matrix with a reasonably chosen ordering is often similar to the order-free matrix. The theoretical results complemented by simulation exercises reveal that the choice of ordering does not significantly affect the results. Simulation experiments also show that when the spatial correlation is weak or moderate, the DAGAR model outperforms CAR models in their ability to correctly estimate a latent spatial surface while the performances are similar for data with stronger spatial dependence. 

\section{Model}\label{sec:dagmain}
\subsection{Cholesky Factors}\label{sec:chol}
{
	We first review a general approach to modeling Gaussian covariance matrices using sparse Cholesky factors and discuss how this relates to CAR models and general covariance estimation. This helps motivate the subsequent construction of our model in Section \ref{sec:dagone}. }
We assume that the graph of the regions is connected. Disconnected graphs with multiple islands will entail a simple extension with block diagonal covariance structures, where each block represents an island. Let $\calG = (V,E)$ denote the connected graph with the regions as vertices $V$ and edges $E$ between neighbors. We denote the $i$th region simply by $i$ and let $A=(a_{ij})$ denote the adjacency matrix for this undirected graph. To model Cholesky factors we specify distributions of the $w_i$'s as  
\begin{equation}\label{eq:telescope}
w_1 = \eps_1,\;
w_2 = b_{21} w_1 + \eps_2,\; \ldots ,\; w_k = b_{k1} w_1 + \cdots + b_{k,k-1} w_{k-1} + \eps_k,
\end{equation}
where the $\eps_i$'s are independent $N(0,\tau_i)$ errors. Throughout this section and Section \ref{sec:dagone}, we assume $\tau_w$, the scale parameter for the $w_i$'s, is one.  
$B=(b_{ij})$ in (\ref{eq:telescope}) is a strictly lower triangular matrix. 
Let $F$ be the diagonal matrix with $\tau_1,\ldots,\tau_k$ on the diagonal. Then $w \sim N(0, L^T F L)$ where $L=I-B$. 
{
	Switching from the specification in (\ref{eq:sar})
	to a strictly lower triangular matrix $B$ is not restrictive because of the following result.
	
	\begin{theorem}\label{th:intrinsic}
		Let $w \sim N(0,Q)$ where $Q$ is the (possibly singular) precision matrix. Then there exists a permutation matrix $P$, a strictly lower triangular matrix $B$ and a diagonal matrix $F$ with non-negative entries such that $PQP^\top=(I-B)^\top F (I-B)$. 
	\end{theorem}
	
	While this result is trivial if $Q$ is non-singular, for rank deficient $Q$ this relies on the algorithm for obtaining the Cholesky factor. All proofs are presented in the Supplement. Hence, any multivariate normal distribution can be expressed as in (\ref{eq:telescope}) under certain orderings of the areal units. For low rank distributions this will be equivalent to setting some of the $\tau_i$'s to zero. 
	In fact, }switching to the lower triangular $B$ has several advantages. First, $L$ is lower triangular with ones on the diagonal, guaranteeing that $L^T F L$ is positive definite as long as all $\tau_i$'s are positive. 
Also $\det(L^TFL)$ is simply $\prod_{i=1}^n \tau_i$ and the quadratic form $w^TL^TFLw$ can be expressed as $\tau_1w_1^2 \ + \sum_{i=2}^k \tau_i (w_i - \sum_{\{j < i\}} w_j b_{ij})^2$, evaluating which requires $\calO(k+s)$ floating point operations (FLOPs) where $s$ is the sparsity, i.e., the number of non-zero entries of $B$. 
Hence, if $B$ is sparse, the joint density of $w$ can be evaluated in an extremely scalable manner. 

To complete the specification in (\ref{eq:telescope}), we need to fully specify the matrices $B$ and $F$. {
	The parameters $\{b_{ij}\}$ and $\{\tau_i\}$ are identifiable up to a marginal precision parameter because the factorization $(I-B)^\top F (I-B)$ is the $LDL^T$ factorization (a variant of Cholesky decomposition with ones on the diagonal of $L$), which is unique. If multiple  observations have been made for each region, we can estimate $B$ and $F$ without imposing simplifying assumptions. 
	In fact, since a sparse representation of the Cholesky factor $I-B$ is desired, the problem reduces to high-dimensional covariance or precision matrix estimation for which there exists a vast inventory of statistical methods including banding \citep{wu03,bickel08}, tapering \citep{cai10}, thresholding \citep{bickel08b,elk08,roth09} and penalization \citep{mein06,glasso,ling12} among others.}

On the other hand, for large point-referenced spatial datasets, 
\citet{nngp, colnngp, dnngp} 
construct sparse Cholesky factor approximations of the precision matrix from a Mat\'ern covariance function \citep{stein99}. 
These approximations are hence derived from an original joint distribution of the spatial random effects. 

However, most areal datasets lack replication that would permit use of fully data-driven learning methods to estimate $B$ and $F$. Also, unlike well defined Mat\'ern Gaussian processes on continuous domains, there is no well defined covariance matrix on arbitrary graphs 
from which one can derive sparse Cholesky factors. In fact, our goal here is the opposite, that is to construct a multivariate Gaussian distribution on graphs starting from the sparse Cholesky factor. 
Consequently, we will make parametric assumptions that will lead to an interpretable covariance model.  

\subsection{Directed acyclic graph autoregressive models}\label{sec:dagone}
To achieve sparsity, we adopt the strategy of defining neighbor sets $N(i)$ such that $b_{ij}=0$ for all $j \notin N(i)$. 
The choice and size of the neighbor sets for areal datasets can be predicated upon the underlying neighborhood graph $\calG$. For $i > 1$, we define $N(i) = \{j < i, j \sim i\}$. The constraint $j < i$ is necessary to endow $B$ with a lower triangular structure. This reduces (\ref{eq:telescope}) to 
\begin{equation}\label{eq:seqspec}
w_1 = \eps_1, \quad w_i = \sum_{j \in N(i)} w_j b_{ij} + \eps_i, \quad (i=2,\ldots,k)
\end{equation}
This specification is analogous to auto-regressive models for time series. In fact, if $w_i$ denotes the response at time $i$, $N(i)$ includes all time points less than $i$ up to a lag of $r$, and $b_{ij}=b_{i-j}$, then (\ref{eq:seqspec}) simply denotes the autoregressive model of order $r$. In a time-series context, where $i$ and $j$ denote time points, assigning the weights $b_{ij}$ based on the temporal lag seems natural, but for irregular areal datasets, enumeration of the areal units does not have any physical interpretation. In the context of image texture analysis, 
\cite{pomm} used different coefficients for $w_j$ in (\ref{eq:seqspec}) based on the direction of neighbors on a regular lattice, to generate images with a wide range of textures. In general, vertices of irregular graphs based on areal datasets do not share such commonality in terms of spatial orientation of their neighbors.  Hence, it is intuitive to assign equal weights to all the neighbors, i.e., letting $b_{ij}=b_i$ for all $j \in N(i)$. A natural choice would be to let $b_i=1/n_{<i}$ and $\tau_i \propto n_{<i}$ where $n_{<i} = |N(i)|$ denotes the cardinality of the neighbor set. This specification is similar to (\ref{eq: car}) except that we are only using the directed neighbors $N(i)$ instead of all neighbors. Since $n_{<1}=0$, this choice of $b_{ij}$ leads to the conundrum of how to specify $\tau_1$. Either we need to define $\tau_1$ in a manner inconsistent with the definition of $\tau_i$ for $i > 1$ or we define $\tau_1=0$ which yields an improper distribution for $w$. We circumvent this using a more general specification described below that includes the degenerate prior with $b_{i} = 1/n_{<i}$ and $\tau_i \propto n_{<i}$ as a limiting case. 

Let $d_{ij}$ denote the length of the shortest path on $\calG$ between nodes $i$ and $j$. If $\calG$ is a tree, i.e., an acyclic graph, then for any $0 \leq \rho <1$, the matrix with elements $\rho^{d_{ij}}$ is positive definite and can be used to model the covariance of $w$. This extends the $\textsc{AR}(1)$ model for time series to any tree graph \citep{mar}. However, graphs corresponding to areal datasets are rarely acyclic and for loopy graphs such results generally do not hold. A spanning tree of a graph is a subgraph that is a tree and includes all the vertices of the original graph. Spanning trees have been used to iteratively approximate parametric covariance matrices over loopy graphs \citep{suddtreethesis}. Borrowing these ideas, a potential solution would be to define the covariance matrix for $w$ as the AR$(1)$ covariance matrix on a spanning tree of $\calG$. However, for large graphs, strategies for deciding upon the best spanning tree are unclear and computationally expensive. Furthermore, as demonstrated in \citep{suddtreethesis}, ignoring certain edges when pruning $\calG$ to a spanning tree can lead to large errors. Instead, we will use local spanning tree embeddings of small subgraphs of $\calG$ to construct the lower dimensional conditional densities specified in (\ref{eq:seqspec}). This method will not ignore any edge and yet produces a computationally convenient precision matrix. 

Let $\calG_i$ be the subgraph of $\calG$ comprising vertices $\{i\} \cup N(i)$ and the edges among them. We intend to construct the conditional density $w_i \mid w_{N(i)}$ using an embedded  spanning tree $\calT_i$ of $\calG_i$. 
The natural candidate for $\calT_i$ is the tree graph $(\{i\} \cup N(i),\{ i \sim j \mid j \in N(i)\})$ as it contains all edges between $i$ and $N(i)$. 
We specify the conditional density $w_i \mid w_{N(i)}$ using the $\textsc{AR}(1)$ model on $\calT_i$ with parameter $\rho$. To be precise, for any $0 \leq \rho < 1$, an auto-regressive AR(1) covariance matrix with parameter $\rho$ on $\calT_i$ is given by 
	\begin{equation}\label{eq:ar}
	\left(	\begin{array}{ccccc}
	1 & \rho & \rho & \cdots & \rho \\
	\rho & 1 & \rho^2 & \cdots & \rho^2 \\
	\rho & \rho^2 & 1 & \cdots & \rho^2 \\
	\vdots & \vdots & \vdots & \vdots & \vdots \\
	\rho & \rho^2 & \cdots & \rho^2 & 1
	\end{array} \right) = \left( 
	\begin{array}{cc} 
	1 & v_i^\top \\
	v_i & \Sigma_i 
	\end{array} \right) .
	\end{equation} This helps us define $E(w_i \mid w_{N(i)} ) =  v_i^T \Sigma_i^{-1} w_{N(i)} $ and $var(w_i \mid w_{N(i)} ) = 1 - v_i^T \Sigma_i^{-1} v_i$, where $v_i$ is the $n_{<i} \times 1$ vector of covariances between $w_i$ and $w_{N(i)}$,  and $\Sigma_i$ is the $n_{<i} \times n_{<i}$ covariance matrix of $w_{N(i)}$ assuming an AR(1) model on $\calT_i$. From equation (\ref{eq:ar}) it is clear that $v_i = \rho 1$ where $1$ denotes the vector of ones, and $\Sigma_i$ is the matrix with one on the diagonals and $\rho^2$ on the off-diagonals. 
Equating this with (\ref{eq:seqspec}), we have 
\begin{align}\label{eq:btau}
b_{ij} = & \frac \rho{1+(n_{<i}-1)\rho^2} \;(i=2,\ldots,k; \;j \in N(i)), \;
\tau_i = & \frac{1+(n_{<i}-1)\rho^2}{1-\rho^2} \;(i=1,\ldots,k)
\end{align}

The specifications in (\ref{eq:btau}) reveal some desirable intuitive features. First, as discussed earlier, $b_{ij}=b_i$ for all $j \in N(i)$, thereby assigning equal weights to all the directed neighbors. Also, the conditional precision $\tau_i$ for $w_i$ increases with the number of directed neighbors. The formulation of $\calT_i$ also ensures that any edge between $i$ and $j$ is incorporated in the conditional specification of $w_i$ or $w_j$ depending on which comes later in the ordering. So, unlike approximating the entire graph with a spanning tree, the local spanning tree approach ensures that no edge of $\calG$ is ignored. Furthermore, for any $0 \leq \rho < 1$ all $\tau_i$'s are positive thereby ensuring a proper probability distribution $w \sim N(0,L^TFL)$. 
The limiting case of $\rho=1$ is equivalent to the improper prior with $b_i=1/n_{<i}$ and $\tau_i \propto n_{<i}$.

The constructions in (\ref{eq:telescope}) and (\ref{eq:btau}) assume a specific ordering, which we now generalize to any other ordering. Let $\pi =\{\pi (1),\ldots,\pi (n)\}$ be any predetermined ordering of the regions and $\pi ^{-1}$ denote its corresponding inverse permutation. Under this ordering, for any $i \neq \pi (1)$, we define its past observations $w_{<i, \pi }$ as the collection $\{ w_j \mid \pi ^{-1}(j) < \pi ^{-1}(i)\}$ and 
its set of directed neighbors $N_\pi  (i)=\{j  \mid i \sim j,\;  \pi ^{-1}(j) < \pi ^{-1}(i)  \}$.
Let $E_\pi $ denote the collection of directed edges from all members of $N_\pi  (i)$ to $i$ for every $i \neq \pi (1)$. We now have a directed acyclic graph $\calD_\pi  = (V, E_\pi )$. Let $n_{\pi (i)} = |N_\pi  (i)|$. The generalization of  (\ref{eq:btau}) based on $D_\pi $ is:
\begin{align}\label{eq: nar}
\begin{array}{c}
w_i \mid w_{<i,\pi } \sim N\left(\frac \rho{1+(n_{\pi (i)}-1)\rho^2}\sum_{j \in N_\pi (i)} w_j, \frac {1+(n_{\pi (i)}-1)\rho^2}{1-\rho^2} \right)\;,
\end{array}
\end{align}
where for any $i$ that has no directed neighbors under $\pi $, $n_{\pi (i)}=0$ and the conditional mean in (\ref{eq: nar}) is zero. If $w_\pi =(w_{\pi (1)},\ldots,w_{\pi (k)})^T$, and $L_\pi $ and $F_\pi $ denote the analogous matrices corresponding to this ordering $\pi $, then we have 
\begin{equation}\label{eq:onedag}
w_\pi  \sim N(0, L^T_\pi  F_\pi  L_\pi )
\end{equation}
This completes the specification of a new class of covariance models for areal datasets. Since the construction is predicated upon a directed acyclic graph derived from an original graph $\calG$, we refer to this as the directed acyclic graph autoregressive or DAGAR model. 
Since $L_\pi $ is lower triangular with $e_\pi=|E_\pi|$ non-zero sub-diagonal entries and $F_\pi $ is diagonal, for any $\rho$, the determinant of $cov(w_\pi )$ is simply  $(1-\rho^2)^k / \prod _{i=1}^k  (1+(n_{\pi (i)}-1)\rho^2)$ and the likelihood for the model in (\ref{eq:onedag}) can be evaluated using $\calO(k+e)$ FLOPs. This ensures that our model is scalable and can be used to analyze massive areal datasets. 

\subsection{Interpretation of $\rho$}\label{sec:rho}
While scalability of the DAGAR model is an important aspect in the analysis of large spatial datasets, our current emphasis is on offering a class of areal  models with an interpretable parameterization. In this regard, we resolve the issue of a lack of meaningful relationship $\rho$ and spatial correlation in the proper CAR models. We now offer insight about the interpretability of $\rho$ in the DAGAR model for certain special graphs. 

\begin{theorem}\label{th:pathcov} Let $\calT$ denote a tree with vertices $V=\{1,\ldots,k\}$ and $\pi $ denote any ordering such that for any $i \neq \pi (1)$, $n_\pi (i) = 1$. Then the covariance matrix in (\ref{eq:onedag}) defines the autoregressive Gaussian process on $\calT$, i.e., $(L_\pi ^TF_\pi  L_\pi )^{-1}=(\rho^{d_{ij}})$ where $d_{ij}$ denotes the shortest path on $\calT$ between $i$ and $j$.
\end{theorem}

Theorem \ref{th:pathcov} shows that if $\calG$ is acyclic, then, under certain orderings including breadth-first and pre-order tree traversals, this model is equivalent to the stationary AR$(1)$ model on trees with $\rho$ being the correlation between neighboring areas. Here, for any two vertices separated by a distance $d$, the correlation is $\rho^d$. 
	The result also shows that while we require an ordering of the locations to construct the DAGAR model, the resulting matrix in this case is order-free and stationary and simply a function of the underlying undirected graph. We now present a result on interpretation of $\rho$ for an $m \times n$ regular grid graph.
	
	\begin{theorem}\label{th:gridcov} Let $\calG$ denote the $m \times n$ grid graph with vertices $V=\{(i,j) \given i=1,\ldots,m; j=1,\ldots,n\}$ and neighbors to the north, south, east, and west, and let $\pi $ denote any diagonal ordering of the vertices corresponding to non-decreasing or non-increasing order of $i+j$ or $i-j$, then $var(w_{(i,j)})=1$ for all $i$ and $j$, and for any neighboring pair of vertices $(i,j)$ and $(i',j')$, $cov(w_{(i,j)},w_{(i',j')})=\rho$. 
	\end{theorem}
	
	Hence, although for a grid, the DAGAR precision matrix is a function of the ordering, for all orderings of Theorem \ref{th:gridcov}, the model yields unit variances and a correlation of $\rho$ for all neighbor pairs. Hence, $\rho$ is still interpretable. This result for a grid graph is quite promising as graphs arising from areal data, like the grid graph, are loopy. Also note that when a CAR model is specified as $w \sim N(0, \tau_w (D-\rho A))$, it may seem that $1/\tau_w$ is the marginal variance of each spatial random effect. Unfortunately, this is not true as the specification of the precision matrix for the CAR model effectuates a heteroskedastic distribution. Consequently, $\tau_w$ can only be interpreted as a common scale factor for the marginal variances. This is remedied in the DAGAR specification, as we see from Theorems \ref{th:pathcov} and \ref{th:gridcov} that the resulting model $N(0, \tau_w Q(\rho))$, where $Q(\rho)$ is the DAGAR precision matrix, is homoskedastic, and hence $1/\tau_w$ is the marginal variance. So, the DAGAR model ensures interpretability for both $\rho$ and $\tau_w$. We will see that this interpretability empowers the DAGAR model to deliver significantly  improved inference about the spatial parameters in areal data analysis. 
	
	Of course, it is difficult to generalize these theoretical interpretability results for irregular graphs. Hence, we also conducted numerical experiments to corroborate the results in Theorems \ref{th:pathcov} and \ref{th:gridcov}, and also gain insight into the relationship between $\rho$ and the neighbor-pair correlations for the proper CAR model and the DAGAR model using an irregular graph. So we used three different graphs: a simple path graph with $100$ vertices which is analogous to a time-series, a two-dimensional $10 \times 10$ lattice or grid graph with edges between vertically or horizontally adjacent vertices, and the state map of the contiguous United States, where two states are said to have an edge if they share a common geographical boundary.
	
\begin{figure}[t]
	\begin{subfigure}[t]{0.3\textwidth}
		\centering
		\includegraphics[scale=0.35,trim={0cm 0cm 1cm 2cm},clip]{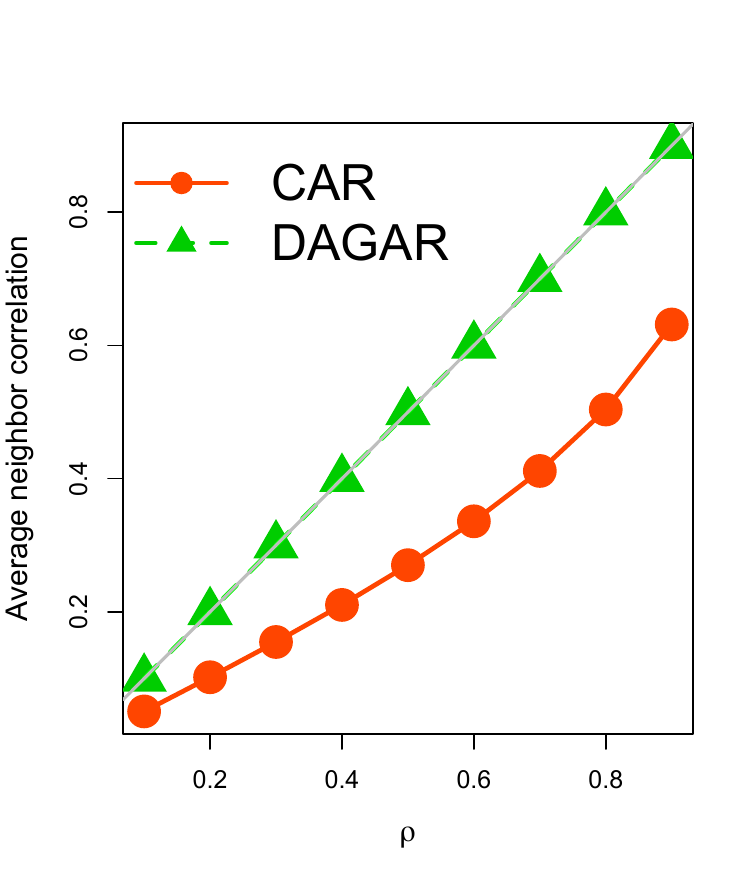}
		\caption{Path graph of length $100$}
	\end{subfigure}
	\begin{subfigure}[t]{0.3\textwidth}
		\centering
				\hskip 4mm \includegraphics[scale=0.35,trim={2cm 0cm 1cm 2cm},clip]{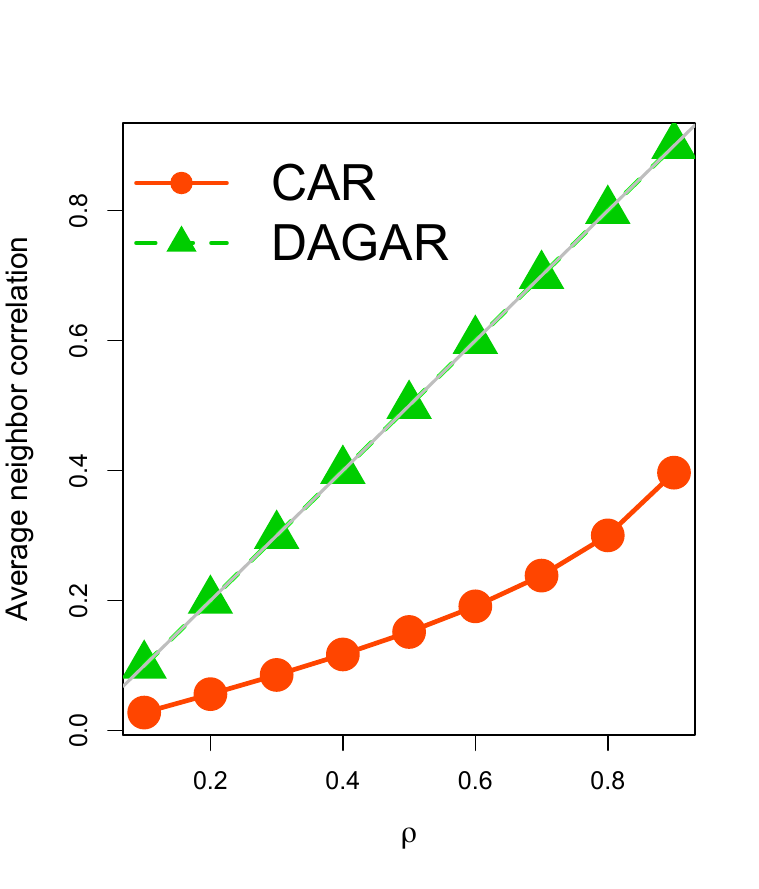}
		\caption{$10\times10$ grid }\label{fig:rhogrid}
	\end{subfigure}
	\begin{subfigure}[t]{0.3\textwidth}
		\centering
		\includegraphics[scale=0.35,trim={2cm 0cm 1cm 2cm},clip]{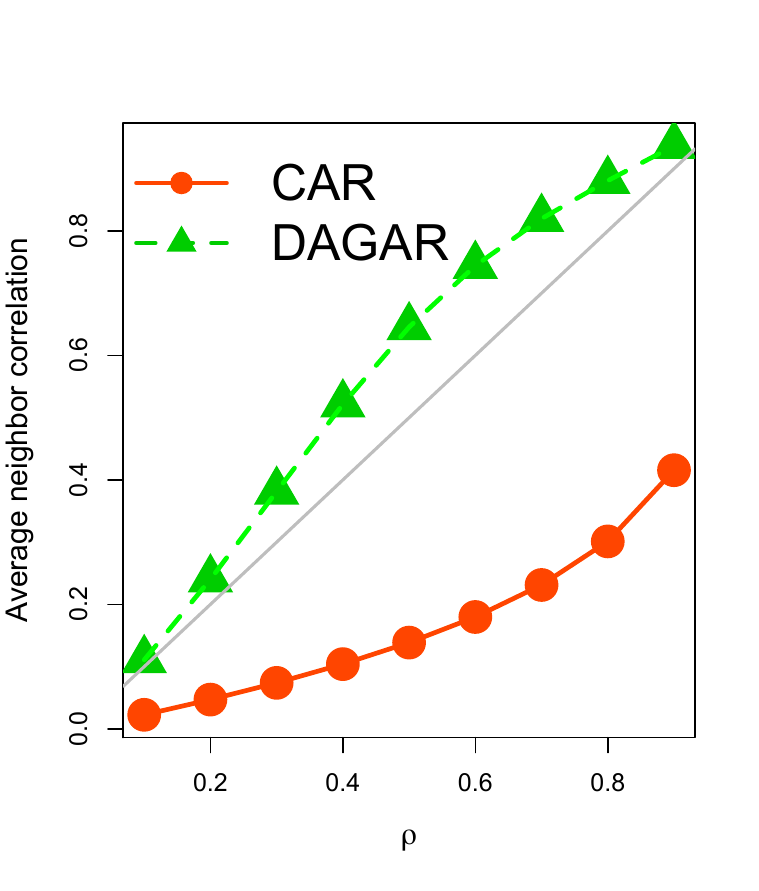}
		\caption{$48$ contiguous US states}\label{fig:rhousa}
	\end{subfigure}
	\caption{Average neighbor pair correlations as a function of $\rho$ for proper CAR and DAGAR model. The solid gray line represents $x=y$ line.}\label{fig:rho}
\end{figure}

We generated covariance matrices corresponding to the two models for $\rho \in \{i/10 \mid i=1,\ldots,9\}$. Figure~\ref{fig:rho} plots the average neighbor-pair correlation, given by $c(\rho)=\sum_{i \sim j} cov(w_i,w_j)/(2\surd var(w_i) \surd var(w_j)) / (\sum n_i)$ as a function of $\rho$, for proper CAR and DAGAR models. For the path and grid graphs, we find the average neighbor pair correlation $c(\rho)$ for our model is exactly $\rho$ as guaranteed in Theorems \ref{th:pathcov} and \ref{th:gridcov}. For the highly irregular United States graph, the theoretical results, of course, do not hold. Nevertheless, $c(\rho)$ for the DAGAR model is much closer to $\rho$ than for the proper CAR model. For the CAR model, even when $\rho$ is close to one, $c(\rho)$ is less than $0.4$. In fact, for all three graphs, the average neighbor-pair correlation for the proper CAR model remains modest. This seems to be true even for very high values of $\rho$ and is consistent with findings elsewhere \citep[see,  e.g.,][]{ban14}. 

\subsection{Impact of Ordering}\label{sec:of}
Unlike covariance or precision matrices that remain invariant up to a permutation factor under different orderings of the multivariate vector, Cholesky factors depend on the ordering of the observations. Our model in (\ref{eq:onedag}) assumes a predetermined ordering $\pi $. We have already seen that for tree and grid graphs, Theorems \ref{th:pathcov} and \ref{th:gridcov} guarantee that the under many different orderings, the DAGAR model retains desirable properties like homoskedasticity and neighbor-pair correlation of $\rho$. 

To understand the impact of ordering beyond the variances and neighbor-pair correlations, we consider an order-free model using a product-of-experts type construction \citep{poe}. Let $P_\pi $ denote the permutation matrix corresponding to $\pi, $ i.e., $P_\pi  (x_1,\ldots,x_k)^T = (x_{\pi (1)},\ldots,x_{\pi (k)} )^T$ for any $k$-dimensional vector $x$, and let $Q$ denote the average of the DAGAR precision matrices in (\ref{eq:onedag}) over all permutations $\pi $, i.e., \begin{equation}\label{eq:dagq}
Q(\rho) = \frac 1{k!} \sum_\pi  P_\pi ^T L^T_\pi  F_\pi  L_\pi  P_\pi \; .
\end{equation}
It is clear that $Q=Q(\rho)$ is free of any ordering and is only a function of the undirected graph $\calG$. Also, since it is the average of positive definite matrices, it is also positive definite. We will use the order-free model $Q$ to understand how the DAGAR precision matrices $Q_\pi $ differ under different choices of the ordering $\pi $. In order to do this, we first note that $Q$ can be expressed in closed form.
\begin{theorem}\label{th:main} Let $i \approx j$ mean that $i$ and $j$ share at least one common neighbor. There exist functions $f(\rho,r)$ and $g(\rho,r)$ for any positive integer $r$ and $0 \leq \rho < 1$ such that 
	\begin{align*}
	Q_{ii} &= 1 + \frac {n_i\rho^2}{2(1-\rho^2)} + \frac {\rho^2}{1-\rho^2} \sum_{j \sim i} f(\rho,n_j) \\
	Q_{ij} &= -\frac{\rho}{1-\rho^2}I(i \sim j) + \frac {1}{1-\rho^2}I(i \approx j) \sum_{k \sim N(i) \cap N(j)} g(\rho,n_k) .
	\end{align*}
\end{theorem}
    Here $I(\cdot)$ denotes the indicator function.
	Explicit expressions for $f(\rho,r)$ and $g(\rho,r)$ are provided in the proof of Theorem~\ref{th:main}. Let $\calK$ denote a set of `reasonable' orderings that one can consider for a given areal dataset. We note that for two orderings $\pi _1$ and $\pi _2$ in $\calK$, the relative difference $\|Q_{\pi _1}(\rho) - Q_{\pi _2}(\rho)\|_F/\|Q(\rho)\|_F$, where $|| \cdot ||_F$ denotes the Frobenius norm, is bounded above by $2 \max_{\pi  \in \calK} \|Q_\pi (\rho) - Q(\rho)\|_F/\|Q(\rho)\|_F$. We now investigate the asymptotic behavior of this quantity for the path and grid graphs. 
		
	\begin{theorem}\label{th:frobpath} Consider the path graph with $k$ nodes, and let $\pi $ denote the left to right or right to left ordering. Then the relative difference
		\begin{equation}\label{eq:frobpath}
		\lim_{k \rightarrow \infty} \frac{||Q_\pi(\rho)-Q(\rho)||_F} {||Q(\rho)||_F} = \sqrt{\frac{4\rho^8 + 2\rho^4}{(3 + 6\rho^2 + \rho^4)^2 + 18\rho^2(1+\rho^2)^2 + 2\rho^4}} .
		\end{equation} 
	\end{theorem}
	
\begin{figure}[t]
		\begin{subfigure}[t]{0.45\textwidth}
			\includegraphics[scale=0.25]{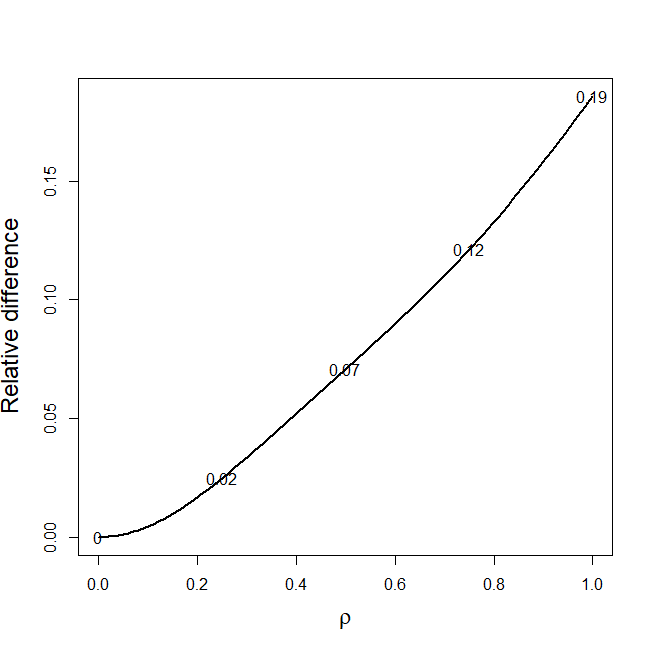}\caption{Path graph}
		\end{subfigure}
		\begin{subfigure}[t]{0.45\textwidth}
			\includegraphics[scale=0.25]{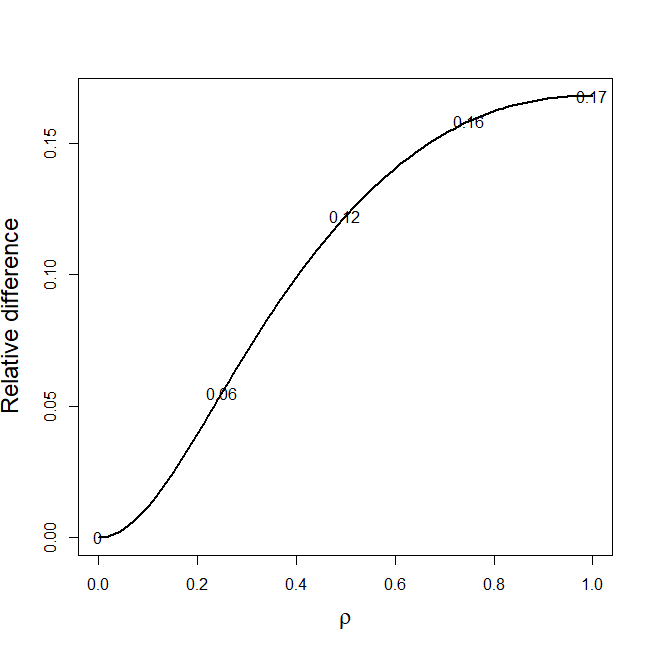}\caption{Grid graph}
		\end{subfigure}
		\caption{Asymptotic relative difference between the DAGAR model and the order free DAGAR model in terms of Frobenius norm for path graph (left) and grid graph (right). The five numbers on each curve corresponds to the values of the difference at $\rho=0,0.25,0.5,0.75$ and $1$ respectively. }\label{fig:frob}
	\end{figure}

	Theorem \ref{th:frobpath} quantifies asymptotically the relative difference between the DAGAR model and the order free version. Figure \ref{fig:frob} plots the quantity on the right hand side of (\ref{eq:frobpath}) as a function of $\rho \in [0,1]$. We see that it is  a monotonically increasing function of $\rho$.  
	For small values of $\rho$ the difference is extremely small ($0.02$ for $\rho=0.25$) and even for moderate values of $\rho$ ($0.5 - 0.75$) the difference is around or less than $10\%$. 
Below we also provide the analogous result for the two-dimensional grid graph. 
	
	\begin{theorem}\label{th:frobgrid} Consider a $m \times m$ grid graph and let $\pi$ denote any of the orderings used in Theorem \ref{th:gridcov}. 
		Then 
		\begin{equation}\label{eq:frobgrid}
		\lim_{m \rightarrow \infty} \frac{||Q_\pi(\rho)-Q(\rho)||_F} {||Q(\rho)||_F} = \sqrt{\frac{\rho^4(\frac {s(\rho)}5-\frac 2 {1+\rho^2})^2+2(\frac 13-\frac{s(\rho)}{30}-\frac{\rho^2}{1+\rho^2})^2+12(\frac 16-\frac{s(\rho)}{60})^2}{(1+\rho^2+\rho^2\frac{s(\rho)}5)^2+4\rho^2+20(\frac 16-\frac{s(\rho)}{60})^2}}
		\end{equation}
		where $s(\rho) = \sum_{r=1}^4 \frac r {1+(r-1)\rho^2}$.
	\end{theorem}
	
	The limit in Theorem \ref{th:frobgrid} looks more complicated than the analogous result for the path graph. However, things are simplified noting that $1-s(\rho)/10$ is $O(\rho^2)$ and, hence, so is the numerator in the right hand side of (\ref{eq:frobgrid}). Figure \ref{fig:frob} (b) plots this quantity as a function of $\rho$. We see that once again this is monotonic in $\rho$ and the difference is negligible for small $\rho$. Theorems \ref{th:frobpath} and \ref{th:frobgrid} show that at least for small $\rho$, the impact of ordering is negligible, though for larger $\rho$ theoretically the DAGAR precision matrices for different orderings will be somewhat different. 

While these results are restricted to the case of special graphs, for an arbitrary areal dataset, one approach would be to use simple intuitive orderings based on the coordinates representing the nodes in some Euclidean embedding of $\calG$. Similar strategies have been used in Cholesky factor based approaches in \cite{nngp}, \cite{stein04} and \cite{ve88} who observed empirically that the joint distribution seemed to be less sensitive to ordering of the regions. Our own set of simulations, detailed in Section \ref{sec:ord} will confirm these finding as we observe that results corresponding to different orderings are similar for both regular and irregular graphs. 

    The order-free model, owing to the availability of the precision matrix in closed form, can be deemed a worthy candidate for analyzing areal data given its liberation from a synthetic ordering. Our simulation analyses detailed in Section~\ref{sec:illu}, show that for a wide range of scenarios, performance of the DAGAR model and its order-free version were very similar. However, the order-free model has certain disadvantages. From Theorem~\ref{th:main}, it is clear that for $i \neq j$, $Q_{ij} \neq 0$ if and only if either $i \sim j$ or $i \approx j$. Hence, the sparsity of $Q$ is $e_2$ where $e_2$ is the number of edges in the second order graph created from $\calG$. As $e_2 > e$, $Q$ is less sparse than the precision matrices for the original DAGAR model in Section~\ref{sec:dagone} or the CAR models. Furthermore, unlike the DAGAR precision matrix $Q_\pi$, $Q$ does not have a closed form expression for the determinant, which invokes expensive computations. These computational roadblocks limit the possibility of using the order-free model for larger datasets. The results of Theorems \ref{th:pathcov} and \ref{th:gridcov} about interpretability of the parameter $\rho$ also do not carry over to the order-free model. 

\section{Data analysis}\label{sec:num}

\subsection{Data generated using an exponential Gaussian Process}\label{sec:illu}
Models for areal datasets are typically used as priors for areal random effects in a hierarchical setup. For example,
let $y_i$ denote the response observed at region $i$ and $x_i$ denote the corresponding set of covariates. A spatial generalized linear mixed model framework assumes $h(E(y_i)) = x^T_i\beta + w_i $ where $h(\cdot)$ denotes a suitable link function. Subsequently, a hierarchical areal model is customarily specified as
\begin{align}\label{eq:bayesmod}
\begin{array}{c}
\prod_{i=1}^k pr(y_i\mid x_i^T\beta+w_i,\theta) \times N(w \mid 0, \tau_w Q(\rho)) \times pr(\beta,\tau_w, \rho, \theta)
\end{array}
\end{align}
where $Q(\rho)$ denotes the precision matrix of the areal model, $pr(y_i\mid x_i^T\beta+w_i, \theta)$ denotes the density corresponding to the link $h(\cdot)$ and $pr(\beta,\tau_w, \rho, \theta)$ is the prior for the parameters. If $h(\cdot)$ is the identity link, e.g., the responses are Gaussian, then we can exploit conjugacy for generating $w$ in a sampler. However, for non-Gaussian responses, we have to sample $w$ and the other parameters using a Metropolis random walk sampler from the joint density in (\ref{eq:bayesmod}).

We conducted simulation experiments assessing the performance of the areal data models using the three graph structures described in Section~\ref{sec:rho} --- path, grid and US states. For each graph, we embed the vertices on the Euclidean plane and generate the spatial random effect vector $w$ from an Gaussian process, i.e., $w \sim N(0,\tau_w M)$ where $M^{-1}$ is the covariance matrix corresponding to an  exponential (Mat\'ern$_{1/2}$) GP, i.e., $M^{-1}=\exp(-\phi d(i,j))$ with $d(i,j)$ denoting the distance between the embedding of the $i^{th}$ and $j^{th}$ vertex. The path graph has a distance preserving embedding in the Euclidean plane such that $D(i,j)=|i-j|$. We embed the grid graph within a $10\times 10$ grid in the Euclidean plane. Although the resulting distance matrix is not identical to the shortest distance (or geodesic) matrix on the graph, the distance between each neighbor-pair remains one. For the United States graph, we use the centroid of each state to create the distance matrix. We scale the distance matrix so that the average neighbor-pair distance is one. To generate $w$, we use $\tau_w=0.25$,  $\tau_e=2.5$ and $\phi=-\log(j/10)$ for $j=1,\ldots,9$. This implies that for the exponential GP the average neighbor pair correlation $\rho=\exp(-\phi)$ varies between $0.1$ and $0.9$, thereby covering a wide spectrum of scenarios. Subsequently, we generate the response $y$ comprising independent $y_i=x_i^T\beta+w_i+\eps_i$, where $x_i$ is a $2\times 1$ vector comprising two independent standard normal variables, $\beta=(1,5)^T$, and the $\eps_i$'s are independent $N(0,\tau_e)$.

We fitted all models using the hierarchical setup in (\ref{eq:bayesmod}) with the six different choices of $Q(\rho)$ --- 1) proper CAR, 2) ICAR, 3) a scaled ICAR model (as suggested by one reviewer) of \cite{sorbye2014scaling} which specifies a prior for $\tau_w$ in the ICAR model as  $\tau_w \sim Gamma(2,\sigma^2_{ref})$ where $$\sigma^2_{ref} = \exp \left( \frac 1k \sum_{i=1}^k \log ((Q^+)_{ii}) \right), $$ $Q^+$ denoting the Moore-Penrose inverse of $Q$, 4) the dimension-reducing sparse GLMM \citep{hughes2013dimension}, and the 5) ordered (DAGAR) and 6) order-free (DAGAR$_{OF}$) directed acyclic graph autoregressive models. To create the directed acyclic graph for our model, we used the ordering based on the sum of the co-ordinates of the mappings of the vertices. The sparse GLMM was implemented directly using the ngSpatial R-package \citep{rngspatial}. For the remaining models, we used conjugate priors $\beta \sim N(0, 0.001 I)$, $\tau_w \sim Gamma(2,1)$ (except for the scaled ICAR for which the prior for $\tau_w$ is specified above) and $\tau_e \sim Gamma(2,0.1)$. For the proper CAR and the two DAGAR models, the spatial correlation parameter $\rho$ was assigned a $Unif(0,1)$ prior. For each combination of parameter values we used $100$ replicate datasets.

\begin{figure}[t!]
	\hskip -0cm\begin{subfigure}[t]{0.4\textwidth}
		\centering
		\includegraphics[scale=0.5,trim={0.6cm 0cm 4.5cm 0cm},clip]{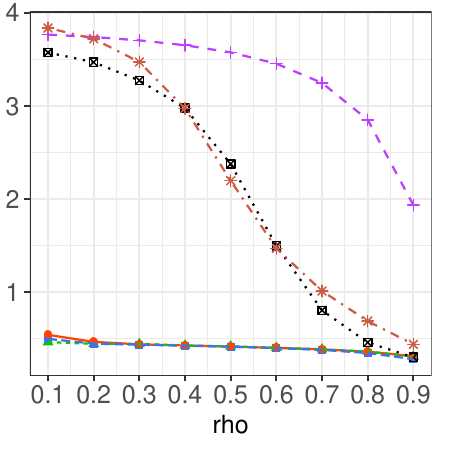}
		\caption{Path}\label{fig:msepath}
	\end{subfigure}
		\hskip -3cm \begin{subfigure}[t]{0.4\textwidth}
	 	\centering
	 	\includegraphics[scale=0.5,trim={0.6cm 0cm 4.3cm 0cm},clip]{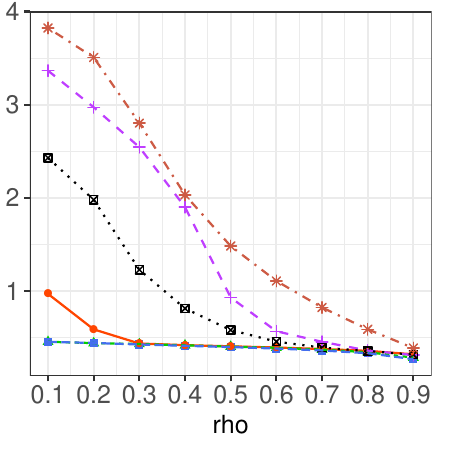}
	 	\caption{Grid}\label{fig:msegrid}
	 \end{subfigure}
     \hskip -2cm \begin{subfigure}[t]{0.4\textwidth}
	 		 	\centering
	 		 	\includegraphics[scale=0.5,trim={0.6cm 0cm 0cm 0cm},clip]{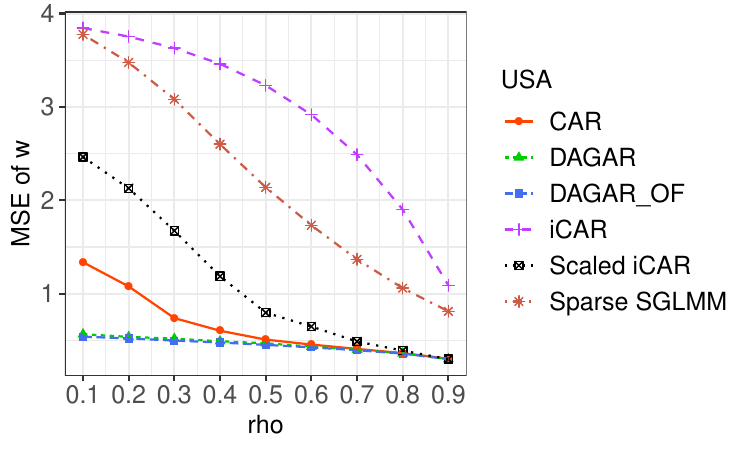}
	 		 	\caption{USA}\label{fig:mseusa}
	 \end{subfigure}
			\caption{MSE as a function of the true $\rho$ (x-axis) for the simulation data analysis using data generated from an exponential GP}\label{fig:gpmse}
\end{figure}

Figure~\ref{fig:gpmse} plots the mean square error (MSE) between the true and estimated $w$ averaged over $100$ replicates for each scenario. We first observe that the mean square errors for the ICAR-based models (ICAR, scaled ICAR and sparse GLMM) are significantly higher than the other three models for all three graphs. The scaled ICAR was the best among these three, producing substantially lower MSEs  than the original ICAR and the sparse GLMM. The sparse GLMM also produced lower MSEs than ICAR for path and USA graphs, but had slightly higher MSE for the grid graph. However, the three models involving the additional $\rho$ parameters, i.e., the proper CAR and the two DAGAR models, consistently produced lower MSEs. For the path graph there is no significant difference in MSE among these three models. However, for the grid and USA graphs when $\rho$ is small, the DAGAR models yielded substantially lower errors. We also noticed that, in terms of MSE, there was very little difference between the performance of the ordered DAGAR model (\ref{eq:onedag}) and its order-free analogue (\ref{eq:dagq}) for most of the scenarios. This result is encouraging as the scalability of the former is a pragmatic solution for analyzing very large areal datasets or networks.

\begin{figure}
	\hskip -0.cm\begin{subfigure}[t]{0.4\textwidth}
		\centering
		\includegraphics[scale=0.5,trim={0.6cm 0cm 3.5cm 0cm},clip]{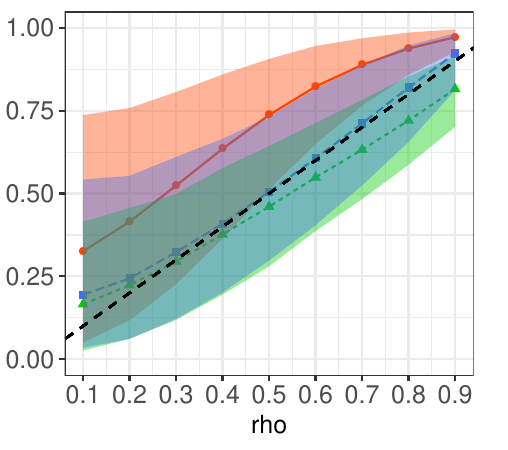}
		\caption{Path}\label{fig:rhoest}
	\end{subfigure} 
		\hskip -2.8cm \begin{subfigure}[t]{0.4\textwidth}
			\centering
			\includegraphics[scale=0.5,trim={1.5cm 0cm 3.5cm 0cm},clip]{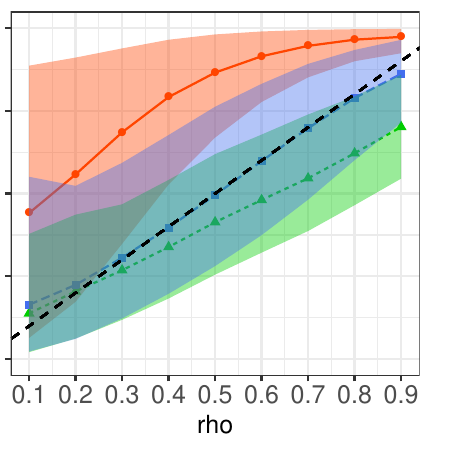}
			\caption{Grid}\label{fig:rhoest}
		\end{subfigure} 
     \hskip -2.3cm \begin{subfigure}[t]{0.4\textwidth}
     	\centering
     	\includegraphics[scale=0.5,trim={1.5cm 0cm 0cm 0cm},clip]{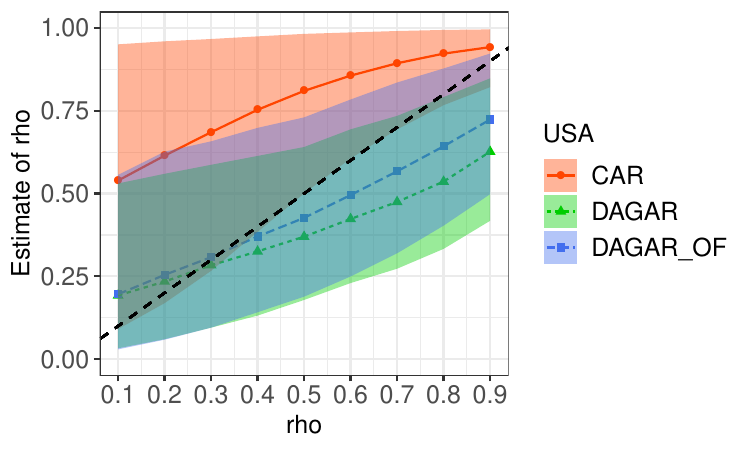}
			\caption{USA}\label{fig:rhoest}
		\end{subfigure}
		\caption{Estimate and confidence bands of $\rho$ as a function of the true $\rho$ (x-axis) for the simulation data analysis using data generated from an exponential GP}\label{fig:rhoest}
	\end{figure}
Next, we consider estimation of $\rho$, as ensuring interpretability of $\rho$ is the motivation driving the construction of the DAGAR model. As the data was generated using an exponential GP, $\rho$ is the unit-distance spatial correlation. We look at the estimates and credible bands of $\rho$ in Figure \ref{fig:rhoest} for the three models involving $\rho$ (the ICAR-based models do not involve $\rho$ and, hence, are not included). We see that for all three graphs, estimates for $\rho$ from the proper CAR model are considerably higher than the true value. The bias is especially stark when the true $\rho$ is small. The DAGAR models generally perform much better in this respect with much less estimation bias, particularly for small $\rho$. For larger $\rho$, the order-free model performs slightly better with the ordered model demonstrating some downwards bias. The $95\%$ confidence bands for both the DAGAR models cover the true value of $\rho$ in most scenarios for all three graphs, while the bands for the proper CAR clearly miss many of the true $\rho$ values. 

\begin{figure}[!]
	\centering
	\hskip -2cm\begin{subfigure}[t]{0.4\textwidth}
		\centering
		\includegraphics[scale=0.5,trim={0.6cm 0cm 4.5cm 0cm},clip]{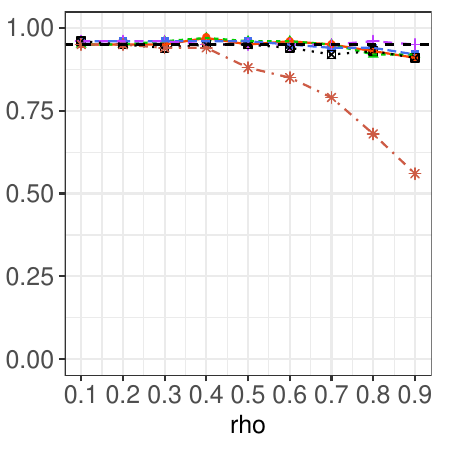}
		\caption{Path: $\beta_1$}\label{fig:cpb1path}
	\end{subfigure}
	\hskip -3cm \begin{subfigure}[t]{0.4\textwidth}
		\centering
		\includegraphics[scale=0.5,trim={0.6cm 0cm 4.5cm 0cm},clip]{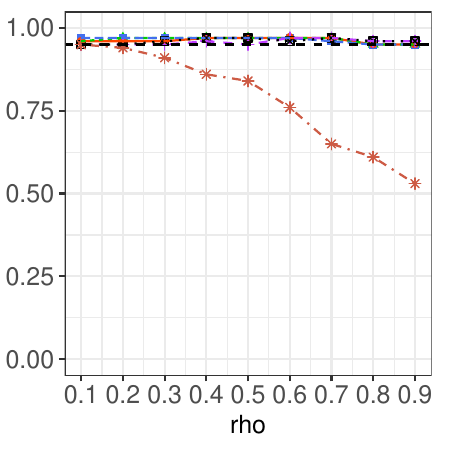}
		\caption{Grid: $\beta_1$}\label{fig:cpb1grid}
	\end{subfigure}
	\hskip -2cm \begin{subfigure}[t]{0.4\textwidth}
		\centering
		\includegraphics[scale=0.5,trim={0.6cm 0cm 0cm 0cm},clip]{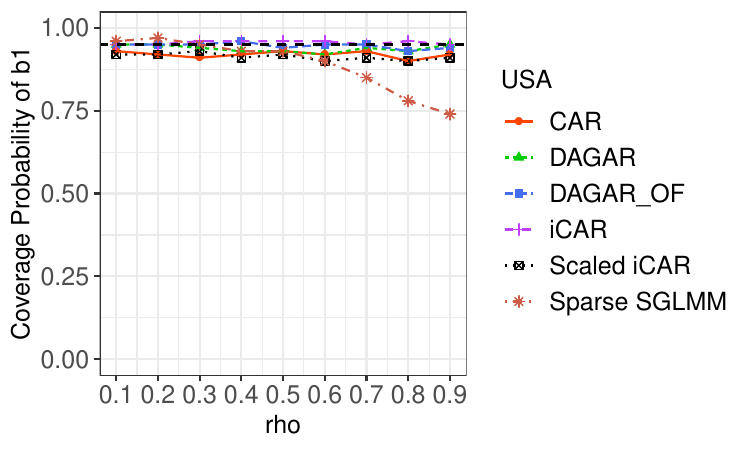}
		\caption{USA: $\beta_1$}\label{fig:cpb1usa}
	\end{subfigure}\\
	\hskip -2cm\begin{subfigure}[t]{0.4\textwidth}
		\centering
		\includegraphics[scale=0.5,trim={0.6cm 0cm 4.5cm 0cm},clip]{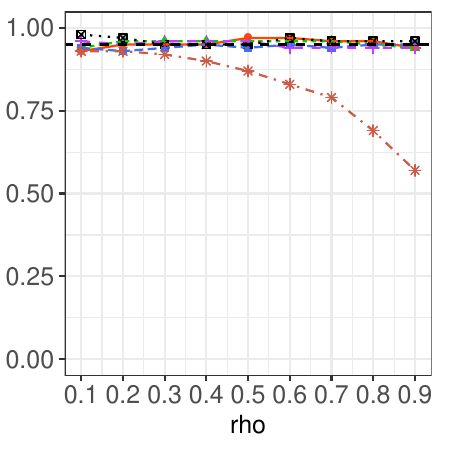}
		\caption{Path: $\beta_2$}\label{fig:cpb2path}
	\end{subfigure}
	\hskip -3cm \begin{subfigure}[t]{0.4\textwidth}
		\centering
		\includegraphics[scale=0.5,trim={0.6cm 0cm 4.5cm 0cm},clip]{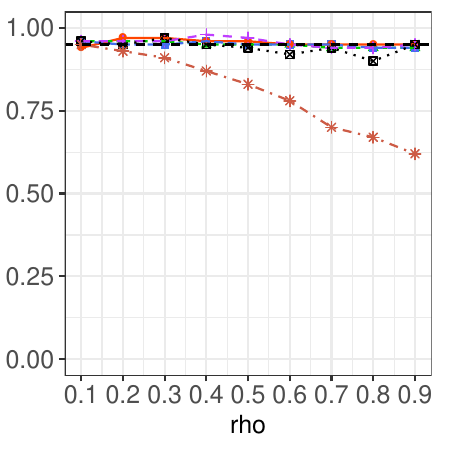}
		\caption{Grid: $\beta_2$}\label{fig:cpb2grid}
	\end{subfigure}
	\hskip -2cm \begin{subfigure}[t]{0.4\textwidth}
		\centering
		\includegraphics[scale=0.5,trim={0.6cm 0cm 0cm 0cm},clip]{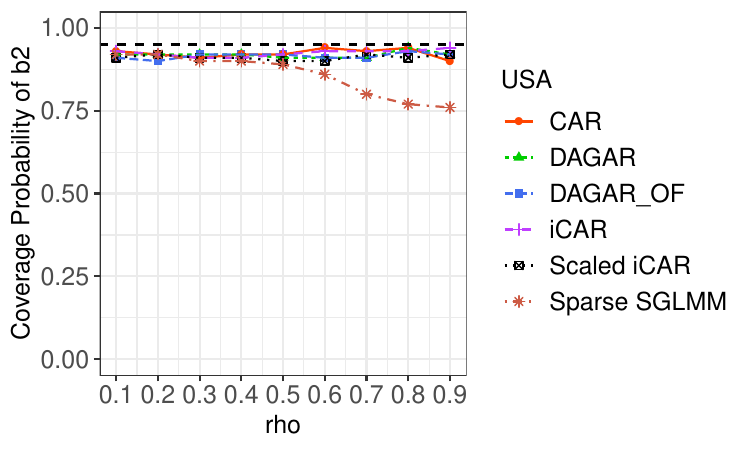}
		\caption{USA: $\beta_2$}\label{fig:cpb2usa}
	\end{subfigure}\\
	\hskip -2cm\begin{subfigure}[t]{0.4\textwidth}
		\centering
		\includegraphics[scale=0.5,trim={0.6cm 0cm 4.5cm 0cm},clip]{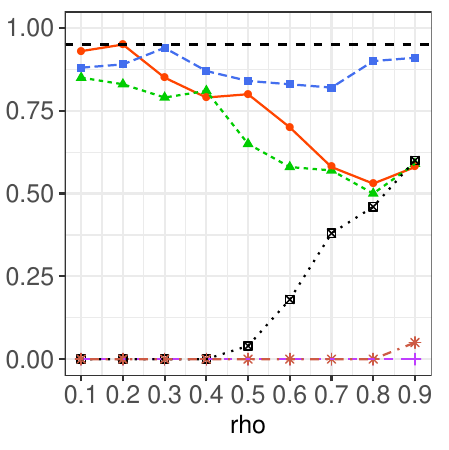}
		\caption{Path: $\sigma^2_e$}\label{fig:cpvarepath}
	\end{subfigure}
	\hskip -3cm \begin{subfigure}[t]{0.4\textwidth}
		\centering
		\includegraphics[scale=0.5,trim={0.6cm 0cm 4.5cm 0cm},clip]{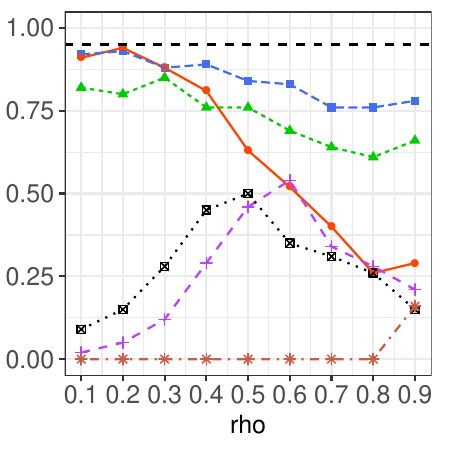}
		\caption{Grid: $\sigma^2_e$}\label{fig:cpvaregrid}
	\end{subfigure}
	\hskip -2cm \begin{subfigure}[t]{0.4\textwidth}
		\centering
		\includegraphics[scale=0.5,trim={0.6cm 0cm 0cm 0cm},clip]{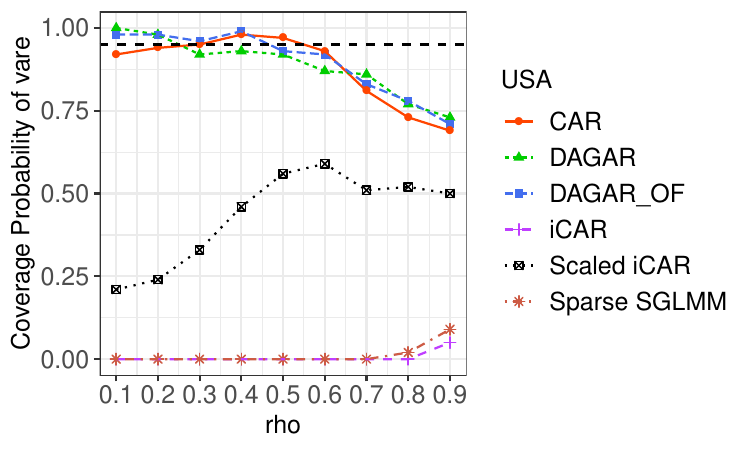}
		\caption{USA: $\sigma^2_e$}\label{fig:cpvareusa}
	\end{subfigure}\\
	\hskip -2cm\begin{subfigure}[t]{0.4\textwidth}
	\centering
	\includegraphics[scale=0.45,trim={0.7cm 0cm 4cm 0cm},clip]{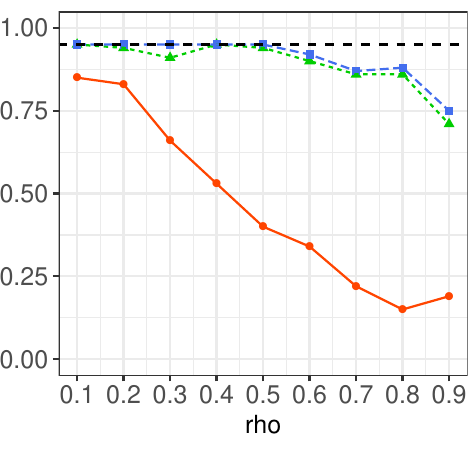}
	\caption{Path: $\rho$}\label{fig:cprhopath}
	\end{subfigure}
	\hskip -3cm \begin{subfigure}[t]{0.4\textwidth}
	\centering
	\includegraphics[scale=0.45,trim={0.7cm 0cm 4cm 0cm},clip]{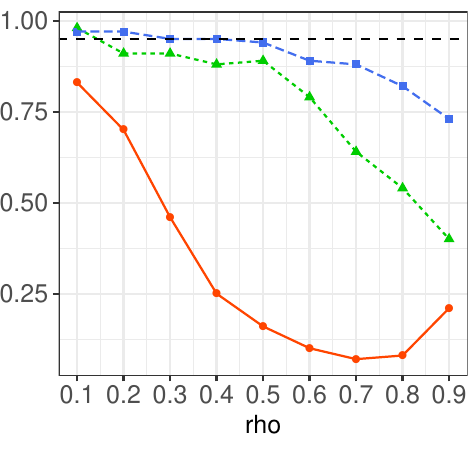}
	\caption{Grid: $\rho$}\label{fig:cprhogrid}
	\end{subfigure}
	\hskip -2cm \begin{subfigure}[t]{0.4\textwidth}
	\centering
	\includegraphics[scale=0.45,trim={0.7cm 0cm 0cm 0cm},clip]{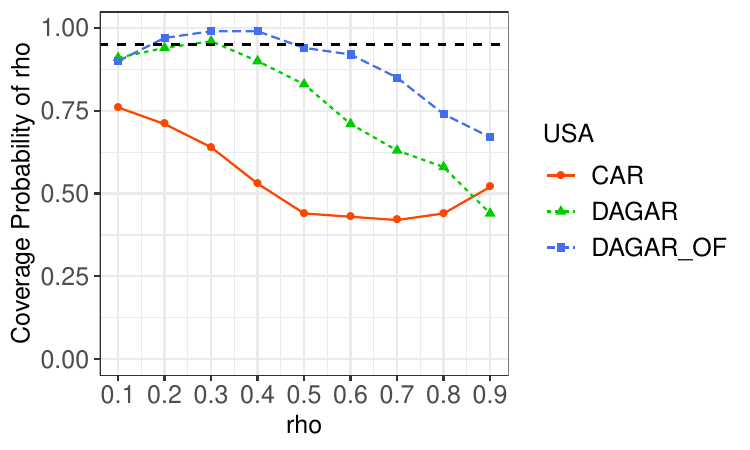}
	\caption{USA: $\rho$}\label{fig:cprhousa}
\end{subfigure}
	\caption{Coverage probabilities of the parameters as a function of the true $\rho$ (x-axis) for the simulation data analysis using data generated from an exponential GP}\label{fig:gpcp}
\end{figure}

Finally, in Figure \ref{fig:gpcp} we plot the coverage probabilities (CP) defined as the mean coverage for a parameter by the $95\%$ confidence intervals over the 100 replications. The regression coefficients $\beta_1$, $\beta_2$ and the error variance $\sigma^2_e=1/\tau_e$ are common to all the models, and hence we compare the CPs for all six models for these parameters. For $\rho$, we only compare the three models using $\rho$. We do not compare $\sigma^2_w=1/\tau_w$, as it has different interpretation for different models. For example, it is the homoskedastic spatial variance for the ordered DAGAR model, but simply a scale factor for the heteroskedastic spatial variances in the CAR models. We see that for $\rho$, the two DAGAR models offer significantly improved coverage over the proper CAR model. This trend was already reflected in the confidence bands in Figure \ref{fig:rhoest} and is confirmed here. For smaller $\rho$ the coverages of the two DAGAR models are nearly identical to the nominal level of $95\%$. While the coverages decline for larger values of $\rho$, they are still almost uniformly and substantially better than the coverage provided by the proper CAR model.  

For the regression coefficients, $\beta_1$ and $\beta_2$, we see that all models except the sparse GLMM offered satisfactory coverage, close to $95\%$. This is not surprising as estimates of regression coefficients are typically robust to variance misspecification. The under coverage of the sparse GLMM is also expected as it tries to adjust for spatial confounding based on an underlying model assumption, which can lead to worse estimates if that assumption is violated. 
{If we generated data in a way such that the eigenvectors correspon to non-zero eigenvalues of the covariance matrix are uncorrelated with the covariates, then it is likely that the sparse GLMM will produce the most accurate estimates as the DAGAR does not adjust for spatial confounding.}  
However, our focus in this manuscript is not on spatial confounding and the data generation scheme we used is extremely common for geo-spatial settings. The dimension reduction approach used by sparse GLMM with the ICAR model as their baseline, can possibly also be adopted for the DAGAR models, to yield versions that guard against confounding. However, care has to be taken avoid eigen decompositions of the covariance matrix at every step of the MCMC, as the DAGAR models, unlike the ICAR, involve $\rho$ whose value will be updated at every iteration. We identify this as one of the future research directions. 

Turning to the error variance $\sigma^2_e=1/\tau_e$, we first note that the ICAR-based models performs surprisingly poorly offering almost zero coverage for small $\rho$ for all the three graphs. Only the scaled ICAR offers somewhat improved coverage for larger values of $\rho$. The CPs from the DAGAR models are once again close to $95\%$ for smaller values of $\rho$ but decline for larger $\rho$. The proper CAR generally offers coverage worse than the DAGAR model and better than the ICAR-based models.

Reliable estimation and inference for spatial covariance parameters is a notoriously difficult problem. These results for the coverage probabilities of the spatial parameters clearly demonstrate the value of our interpretable model in delivering more accurate inference about the parameters for areal data. The results present strong evidence for the superiority of the DAGAR model both in terms of effectively recovering the latent spatial surface and the ability to assess hypotheses related to parameters describing the spatial structure in the data. 

\subsection{Analyses using different orderings}\label{sec:ord} The DAGAR model used in the analyses in Section \ref{sec:illu} for the USA graph was constructed by ordering the nodes (states) from the southwest to the northeast. In this section, we repeat the analysis for the DAGAR model using three other orderings which start at southeast, northwest and northeast respectively and go approximately diagonally to the opposite end of the map. 

\begin{figure}[t!]
		\begin{adjustwidth}{-0cm}{-0cm}
	\begin{subfigure}[t]{0.4\textwidth}
		\centering
		\includegraphics[scale=0.6,trim={0cm 0cm 0cm 0cm},clip]{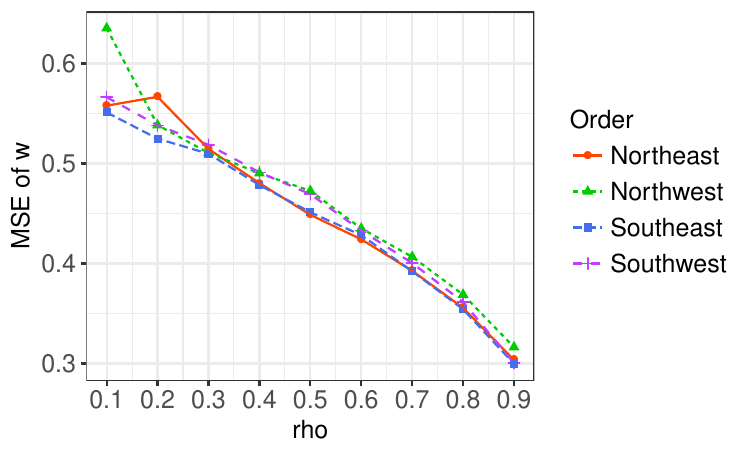}
		\caption{USA: MSE}\label{fig:permmse}
	\end{subfigure}
	\hskip 1cm \begin{subfigure}[t]{0.4\textwidth}
		\centering
		\includegraphics[scale=0.6,trim={0cm 0cm 0cm 0cm},clip]{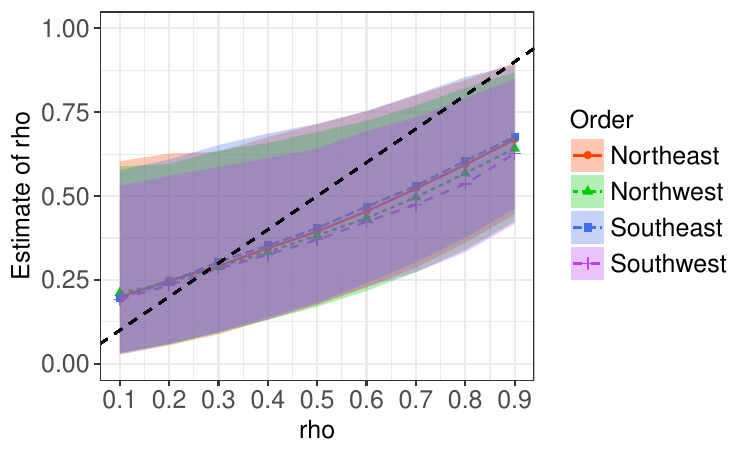}
		\caption{USA: Estimate and confidence bands of $\rho$}\label{fig:permrho}
	\end{subfigure} 
	\caption{MSE (left) and estimates and confidence bands of $\rho$ (right) as a function of the true $\rho$ (x-axis) for four different orderings of the DAGAR model}\label{fig:order}
	\end{adjustwidth}
\end{figure}

Figure \ref{fig:order} plots the average MSE (left) and the estimates and confidence bands for $\rho$ over 100 replicated datasets for the DAGAR model using these three orderings and the original ordering used in Section \ref{sec:illu}. We see that the ordering has little impact on the results as the MSE as well as the estimates and confidence bands for $\rho$ for the four different orderings are nearly indistinguishable.

\subsection{Data generated using CAR and DAGAR}\label{sec:cardagar}
In Sections \ref{sec:illu} and \ref{sec:mgp}, we generated the data using Gaussian Processes to ensure that the data generating mechanism is different from all the models we are fitting to the data (except for the path graph for data generated using an exponential GP). In this Section, we considered simulation schemes where the data were generated using the DAGAR or the proper CAR model for all three graphs. All the parameter choices were kept the same as in Section \ref{sec:illu} and 100 replicates were used for each setting. 

\begin{figure}[h!]
	\hskip -2cm\begin{subfigure}[t]{0.4\textwidth}
		\centering
		\includegraphics[scale=0.5,trim={0.6cm 0cm 4.5cm 0cm},clip]{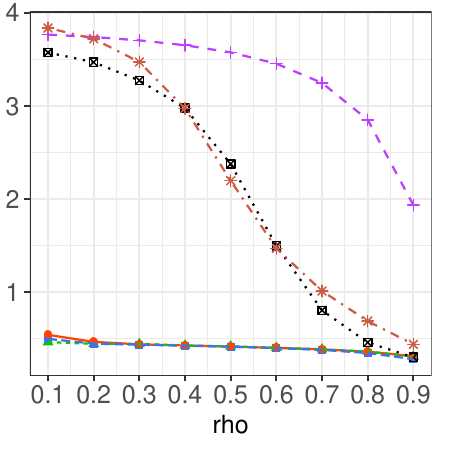}
		\caption{Path}\label{fig:msepath}
	\end{subfigure}
	\hskip -3cm \begin{subfigure}[t]{0.4\textwidth}
		\centering
		\includegraphics[scale=0.5,trim={0.6cm 0cm 4.3cm 0cm},clip]{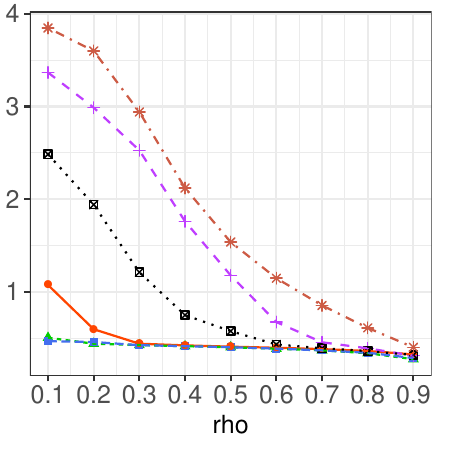}
		\caption{Grid}\label{fig:msegrid}
	\end{subfigure}
	\hskip -2cm \begin{subfigure}[t]{0.4\textwidth}
		\centering
		\includegraphics[scale=0.5,trim={0.6cm 0cm 0cm 0cm},clip]{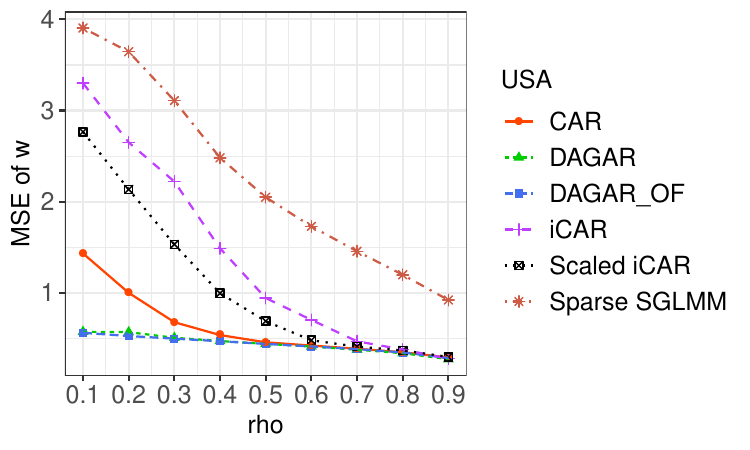}
		\caption{USA}\label{fig:mseusa}
	\end{subfigure}
	\caption{MSE as a function of the true $\rho$ (x-axis) for the simulation data analysis using data generated from a DAGAR model}\label{fig:dagar}
\end{figure}

\begin{figure}[h]
	\hskip -2cm\begin{subfigure}[t]{0.4\textwidth}
		\centering
		\includegraphics[scale=0.5,trim={0.6cm 0cm 4.5cm 0cm},clip]{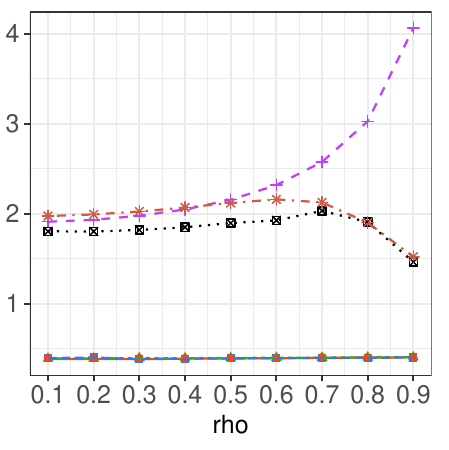}
		\caption{Path}\label{fig:msepath}
	\end{subfigure}
	\hskip -3cm \begin{subfigure}[t]{0.4\textwidth}
		\centering
		\includegraphics[scale=0.5,trim={0.6cm 0cm 4.3cm 0cm},clip]{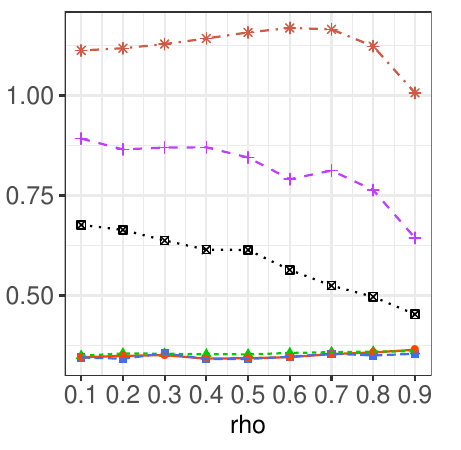}
		\caption{Grid}\label{fig:msegrid}
	\end{subfigure}
	\hskip -2cm \begin{subfigure}[t]{0.4\textwidth}
		\centering
		\includegraphics[scale=0.5,trim={0.6cm 0cm 0cm 0cm},clip]{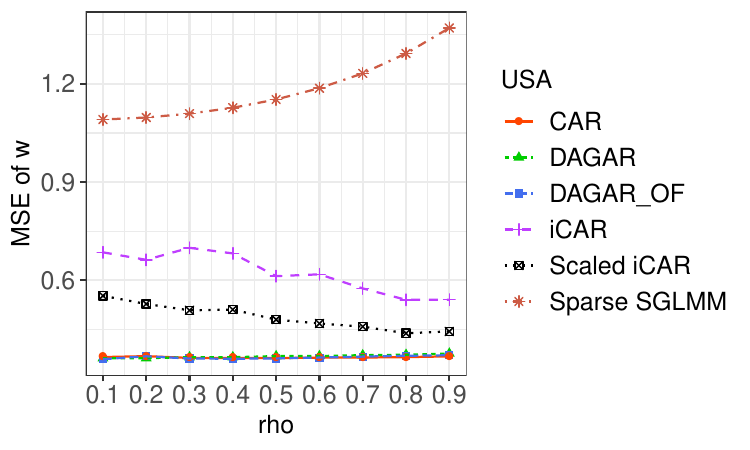}
		\caption{USA}\label{fig:mseusa}
	\end{subfigure}
	\caption{MSE as a function of the true $\rho$ (x-axis) for the simulation data analysis using data generated from a proper CAR model}\label{fig:car}
\end{figure}

Figures \ref{fig:dagar} and \ref{fig:car} plot the average MSE when the data is generated using a DAGAR model. We see that when the data is generated using a DAGAR covariance, the DAGAR model substantially outperforms all the ICAR-based models with significantly lower MSE for all three graphs and all values of $\rho$. For the path graph, the proper CAR also produces MSEs similar to the DAGAR model, whereas for grid and USA graphs for smaller values of $\rho$, the MSE is higher than those of the DAGAR models. The trends in MSE are broadly similar to what was observed in Sections \ref{sec:illu} and \ref{sec:mgp}. When data is generated using a proper CAR, Figure \ref{fig:car} reveals that the DAGAR models, alongwith the proper CAR model (which is the true model), once again produce MSEs substantially lower than the ICAR-based models for all 3 graphs. 

\subsection{Additional analyses}\label{sec:add}
The Supplemental file contains additional analyses for a) simulated data generated with spatial random effects coming from a smoother Mat\'ern$_{3/2}$ GP instead of the exponential GP  (Section \ref{sec:mgp}), and b) non-Gaussian (Poisson) areal data (Section \ref{sec:ng}). Overall, the findings from these analysis concur with the results presented here. Across all the simulation scenarios, we found the performance of the DAGAR model to be remarkably robust, uniformly producing the lowest MSEs, accurately estimating the regression coefficients, error variances, and most remarkably the spatial correlation $\rho$, which is considered to be notoriously difficult to estimate. With the exception of data  generated on a path graph using an exponential GP, all the other settings effectively correspond to misspecified models, and the estimates of the regression coefficients from the DAGAR model were quite robust to this. 
The proper CAR model also performed quite well, often producing MSEs close to those from the DAGAR models except for cases when the true spatial correlation was weak, i.e., $\rho$ was small. However, both estimation and inference for $\rho$ from the proper CAR model was generally much less accurate than the DAGAR model. We note that across all scenarios, the ICAR-based models generally performed quite poorly, producing large MSE, with the scaled ICAR being the best in this class. Finally, the order-free DAGAR model produced results very similar to the ordered DAGAR model for all scenarios. However, it was much slower, especially due to determinant calculations. While using state-of-the-art sparse matrix algorithms would definitely help scale reduce the computing times for the order-free model, that is not the focus of the current manuscript and hence, we do not consider the order-free model for the real data analyses. 

\section{County-level US infant mortality data}\label{sec:inf} We now analyze a large areal data using the DAGAR model. The dataset consists of counts of infant births $B_i$ and deaths $D_i$ for each of $3071$ US counties. County-specific covariates, which possibly affect infant death rates, were available and include number of births with low weight (low$_i$), percentages of black residents (black$_i$) and Hispanic residents (Hisp$_i$), a Gini index measuring income disparity (gini$_i$), social affluence (aff$_i$), and a measure of residential stability (stab$_i$). The dataset is publicly available in the `ngspatial' package in R and was analyzed in \cite{hughes2013dimension} where more description of the data is available. 

We analyzed this dataset using Poisson spatial regression model where each $D_i$ is modeled as an independent Poisson random variable with mean $B_i \exp(\alpha+\beta_1 \mbox{ low}_i + \beta_2 \mbox{ black}_i + \beta_3 \mbox{ Hisp}_i + \beta_4 \mbox{ gini}_i + \beta_5 \mbox{ aff}_i + \beta_6 \mbox{ stab}_i + w_i)$ 
where the $w_i$'s are the spatial random effects.  We assign $\alpha$ and the $\beta_i$s independent $N(0,10^{-4})$ priors. We present the results for $w \sim N(0,\tau_w Q)$ where $Q$ is either the DAGAR or the ICAR model. We could not implement the proper CAR model for such a large dataset. However, we also add the results of the sparse spatial GLMM model. 
In addition to presenting the parameter estimates and confidence intervals, we also use model comparison metrics to evaluate the three covariance models. We used the Deviance Information Criterion \citep[DIC,][]{spieg02} to compare the posterior distributions. Table \ref{tab:inf} presents the results for the three models. Among the seven regression coefficients, estimates for six of them were similar across the three models, with each of the credible intervals yielding the same inference. The exception to this was $\beta_4$, whose estimates differed substantially between the sparse GLMM and the other two models. The sparse GLMM was the only one yielding a credible interval that does not cover zero. 

We do not know if the difference for $\beta_4$ was due to the sparse GLMM accounting for spatial confounding, as this can only be answered depending on what we believe the true data generation process was. We have seen consistently in the simulation analyses using the usual data generation paradigm, how the sparse GLMM offered higher MSEs and poor inference on the regression coefficients. Also, while the credible intervals for the sparse GLMM were significant for all 7 regression coefficients (compared to 6 for the other two models), it also produced a slightly higher DIC than the ICAR model despite being a dimension reduction approach with a fewer number of parameters. 
\begin{table}[!h]
	\caption{Parameter estimates (posterior medians) and model comparison metrics for the US infant mortality data. The numbers inside braces indicates the lower and upper bounds for the $95\%$ credible intervals}\label{tab:inf}
	\begin{adjustwidth}{-5mm}{-1cm}
		\begin{small}
			\begin{tabular}{cccc}
				&&&\\
				& DAGAR & ICAR & sparse SGLMM \\  \hline 
				$\alpha$ & -5.623	(-5.944, -5.353) & -5.641	(-5.871, -5.413) & -5.430	(-5.616, -5.246) \\
				$\beta_1$ & 7.803	(6.438, 9.172) & 7.716	(3.924 9.166) & 8.777	(7.540, 10.032) \\
				$\beta_2$ & 0.00376	(0.00208, 0.00543) & 0.00364	(0.00182, 0.00915) & 0.00423	(0.00288, 0.00556)\\
				$\beta_3$ & -0.00347	(-0.00501, -0.00189) & -0.00286	(-0.00859, -0.00262) & -0.00379	(-0.00488, -0.00270)\\
				$\beta_4$ & -0.0616	(-0.570, 0.480) & 0.103	(-0.425, 0.631) & -0.555	(-0.977, -0.125)\\
				$\beta_5$ & -0.0770	(-0.0911, -0.0632) & -0.0778	(-0.0935, -0.0616) & -0.0757	(-0.0877, -0.0638) \\
				$\beta_6$ & -0.0413	(-0.0590, -0.0234) & -0.0448	(-0.0643, -0.0249) & -0.0285	(-0.0433, -0.0138)\\
				$\tau_w$ & 7.544	(3.615, 12.866) & 32.080	(14.11, 39.87) & 9.450	(3.870, 16.459)\\
				$\rho$ & 0.987	(0.974, 0.995) & & \\
				DIC & 10145.8 & 9902.0 & 10110.0
			\end{tabular}
		\end{small}
	\end{adjustwidth}
\end{table}

DAGAR was the only model to accommodate $\rho$, and the estimate and confidence intervals suggest strong spatial correlation. We have seen consistently from the simulation exercises that when the underlying spatial correlation is strong, the DAGAR model performs similarly to the CAR models. This is consistent with what we observe here. In fact, the deviance information criterion of the three models are within approximately $1\%$ of each other, demonstrating the competitive performance of the DAGAR model even for large datasets in a non-Gaussian setup under strong spatial dependence. Moreover, through the estimation of $\rho$, DAGAR provides additional insight about the spatial dependence that is not offered by the ICAR model or the sparse SGLMM model.

\section{Discussion}\label{sec:conc}
	The existing repertoire of covariance models used for analyzing areal datasets is extremely limited. In this manuscript, we have developed an alternative parametric model for areal datasets that promises to be a significant addition to this inventory. The parametric DAGAR models we have proposed in Section \ref{sec:dagmain}
	are novel and offer a greater degree of interpretability than the CAR models, and will be scalable for large datasets. We observe that when spatial dependence is weak or modest, the DAGAR model excels over both variants of the CAR model, while the results are similar when there is strong spatial correlation. Since, the magnitude of the underlying spatial correlation is unknown apriori in most real life applications, we believe the DAGAR model will be a useful alternative to the CAR models. While the ordering of the locations for the DAGAR model is artificial, the theoretical results and extensive simulations strongly suggest that substantive inference from the DAGAR model is expected to be robust to the ordering.


Analyzing prevalence of many diseases simultaneously in a multivariate setup is becoming increasingly important to accommodate the correlations among different disease prevalences. Many of the popular approaches rely on Cholesky factors of conditionally autoregressive precision matrices \citep{gelmdm03,migmart13,migmart17} which can be computationally prohibitive for large $k$. Our ordered model lends itself naturally to these settings due to the readily available Cholesky decomposition and, hence, promises to broaden the inventory of multivariate disease mapping models. The ordered model also offers a coherent way of modeling on arbitrary graphs or networks of growing size, i.e., if a new point is added to the graph, the nested distributions remain same, unlike  any of the other three models considered here. 




\appendix

\section{Additional simulation analyses}\label{sec:simdet}
\subsection{Data generated using a smoother Gaussian Process}\label{sec:mgp} As pointed out by one reviewer, in the simulation settings of Section \ref{sec:illu}, the data generation model using an exponential GP becomes same as the DAGAR model for the path graph. While this is not true for the grid and USA graphs, and the results were generally consistent across the choice of the graphs, in this section we tried a different data generation model to assess the performance of the areal models. Keeping all other model specifications same, we generated the spatial random effects $w_i$ using a smoother Mat\'ern$_{3/2}$ GP instead of an exponential GP. This ensures that the data generation model does not correspond to any of the six models fitted to the data for any of the three graphs. 

\begin{figure}[h!]
	\begin{adjustwidth}{-0cm}{-0cm}
		\hskip -0.5cm\begin{subfigure}[t]{0.4\textwidth}
			\centering
			\includegraphics[scale=0.5,trim={0.6cm 0cm 4.5cm 0cm},clip]{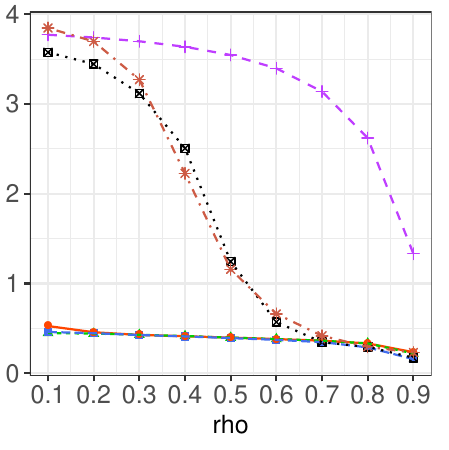}
			\caption{Path}\label{fig:msepath}
		\end{subfigure}
		\hskip -3cm \begin{subfigure}[t]{0.4\textwidth}
			\centering
			\includegraphics[scale=0.5,trim={0.6cm 0cm 4.5cm 0cm},clip]{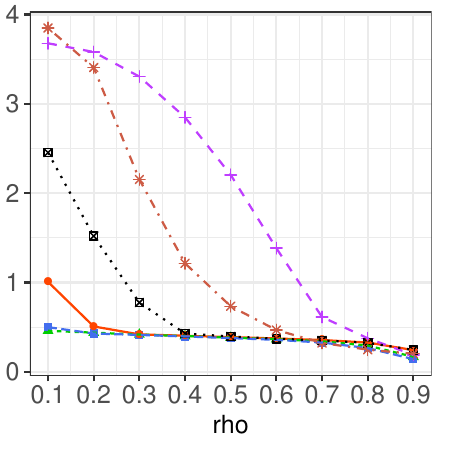}
			\caption{Grid}\label{fig:msegrid}
		\end{subfigure}
		\hskip -2cm \begin{subfigure}[t]{0.4\textwidth}
			\centering
			\includegraphics[scale=0.5,trim={0.6cm 0cm 0cm 0cm},clip]{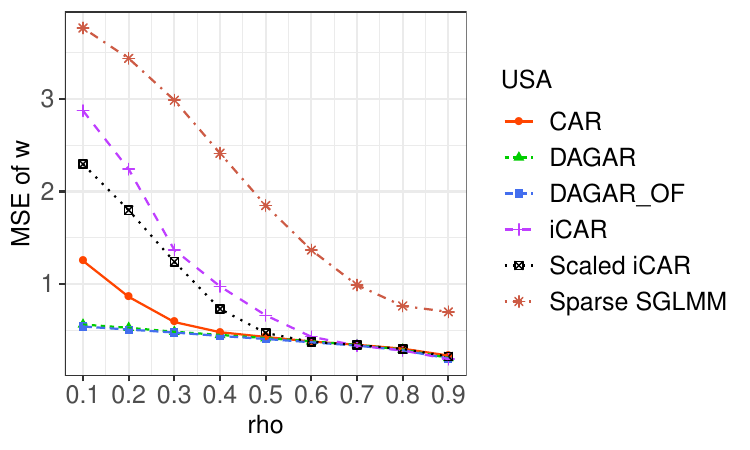}
			\caption{USA}\label{fig:mseusa}
		\end{subfigure}
	\end{adjustwidth}
	\caption{MSE as a function of the true $\rho$ (x-axis) for the simulation data analysis using data generated from a Mat\'ern$_{3/2}$ GP}\label{fig:mgpmse}
\end{figure}

\begin{figure}[h!]
	\centering
	\hskip -2cm\begin{subfigure}[t]{0.4\textwidth}
		\centering
		\includegraphics[scale=0.5,trim={0.6cm 0cm 4.5cm 0cm},clip]{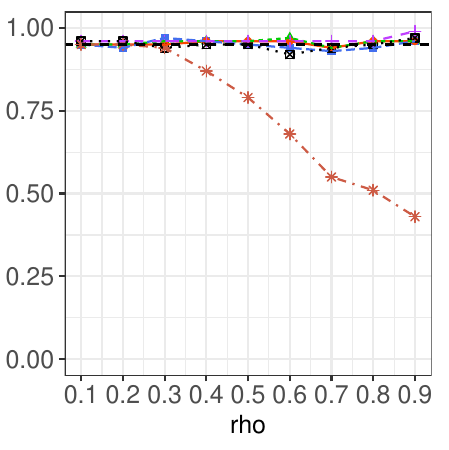}
		\caption{Path: $\beta_1$}\label{fig:cpb1path}
	\end{subfigure}
	\hskip -3cm \begin{subfigure}[t]{0.4\textwidth}
		\centering
		\includegraphics[scale=0.5,trim={0.6cm 0cm 4.5cm 0cm},clip]{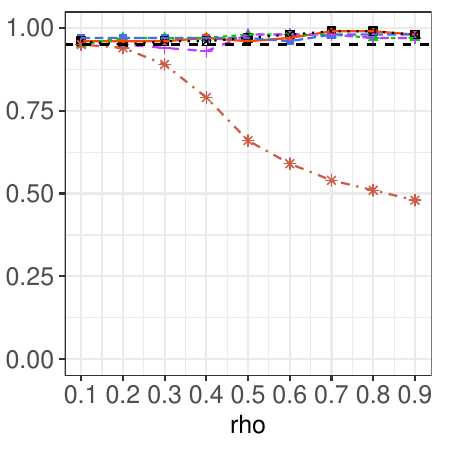}
		\caption{Grid: $\beta_1$}\label{fig:cpb1grid}
	\end{subfigure}
	\hskip -2cm \begin{subfigure}[t]{0.4\textwidth}
		\centering
		\includegraphics[scale=0.5,trim={0.6cm 0cm 0cm 0cm},clip]{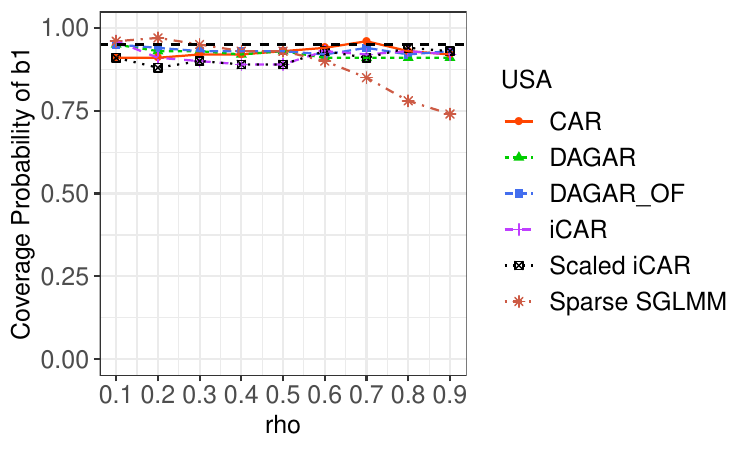}
		\caption{USA: $\beta_1$}\label{fig:cpb1usa}
	\end{subfigure}\\
	\hskip -2cm\begin{subfigure}[t]{0.4\textwidth}
		\centering
		\includegraphics[scale=0.5,trim={0.6cm 0cm 4.5cm 0cm},clip]{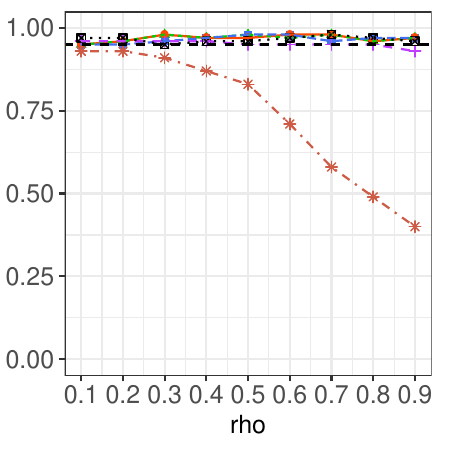}
		\caption{Path: $\beta_2$}\label{fig:cpb2path}
	\end{subfigure}
	\hskip -3cm \begin{subfigure}[t]{0.4\textwidth}
		\centering
		\includegraphics[scale=0.5,trim={0.6cm 0cm 4.5cm 0cm},clip]{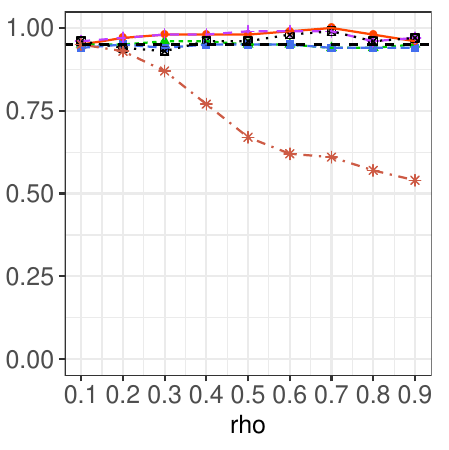}
		\caption{Grid: $\beta_2$}\label{fig:cpb2grid}
	\end{subfigure}
	\hskip -2cm \begin{subfigure}[t]{0.4\textwidth}
		\centering
		\includegraphics[scale=0.5,trim={0.6cm 0cm 0cm 0cm},clip]{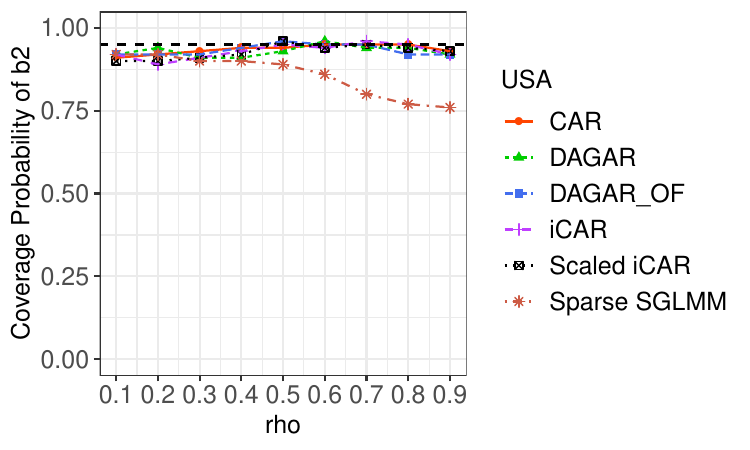}
		\caption{USA: $\beta_2$}\label{fig:cpb2usa}
	\end{subfigure}\\
	\hskip -2cm\begin{subfigure}[t]{0.4\textwidth}
		\centering
		\includegraphics[scale=0.5,trim={0.6cm 0cm 4.5cm 0cm},clip]{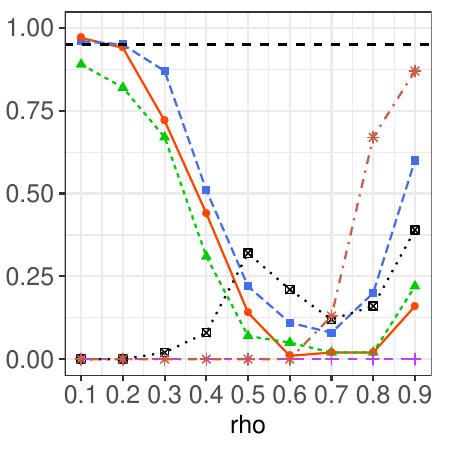}
		\caption{Path: $\sigma^2_e$}\label{fig:cpvarepath}
	\end{subfigure}
	\hskip -3cm \begin{subfigure}[t]{0.4\textwidth}
		\centering
		\includegraphics[scale=0.5,trim={0.6cm 0cm 4.5cm 0cm},clip]{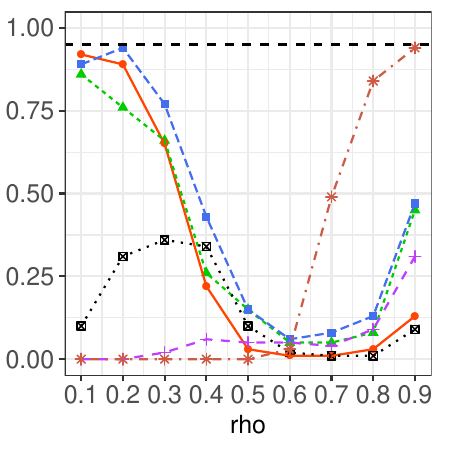}
		\caption{Grid: $\sigma^2_e$}\label{fig:cpvaregrid}
	\end{subfigure}
	\hskip -2cm \begin{subfigure}[t]{0.4\textwidth}
		\centering
		\includegraphics[scale=0.5,trim={0.6cm 0cm 0cm 0cm},clip]{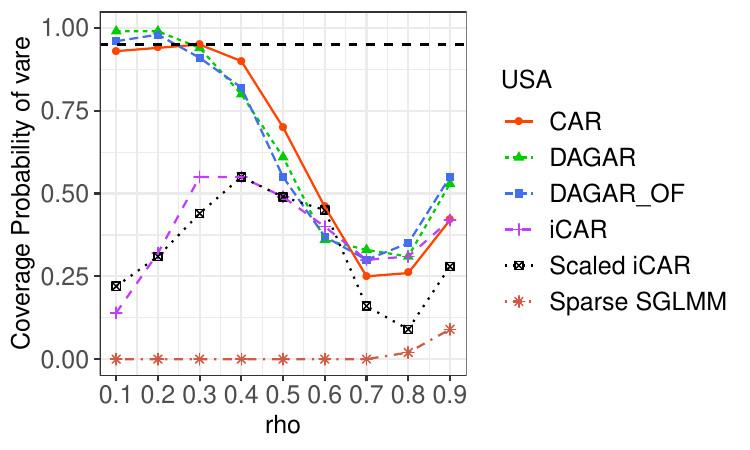}
		\caption{USA: $\sigma^2_e$}\label{fig:cpvareusa}
	\end{subfigure}\\
	\hskip -2cm\begin{subfigure}[t]{0.4\textwidth}
		\centering
		\includegraphics[scale=0.45,trim={0.7cm 0cm 4cm 0cm},clip]{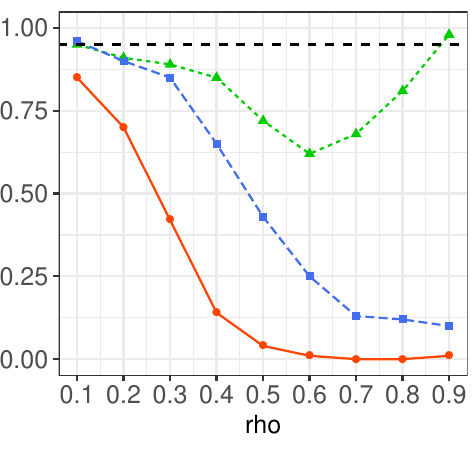}
		\caption{Path: $\rho$}\label{fig:cprhopath}
	\end{subfigure}
	\hskip -3cm \begin{subfigure}[t]{0.4\textwidth}
		\centering
		\includegraphics[scale=0.45,trim={0.7cm 0cm 4cm 0cm},clip]{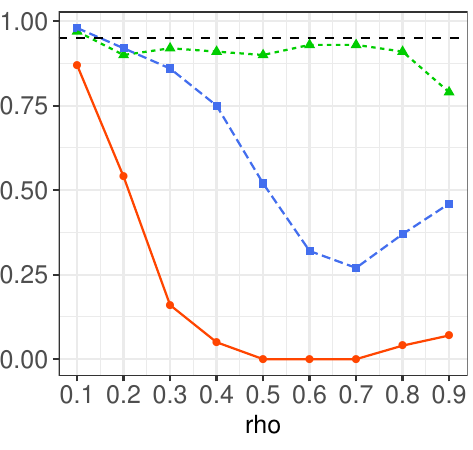}
		\caption{Grid: $\rho$}\label{fig:cprhogrid}
	\end{subfigure}
	\hskip -2cm \begin{subfigure}[t]{0.4\textwidth}
		\centering
		\includegraphics[scale=0.45,trim={0.7cm 0cm 0cm 0cm},clip]{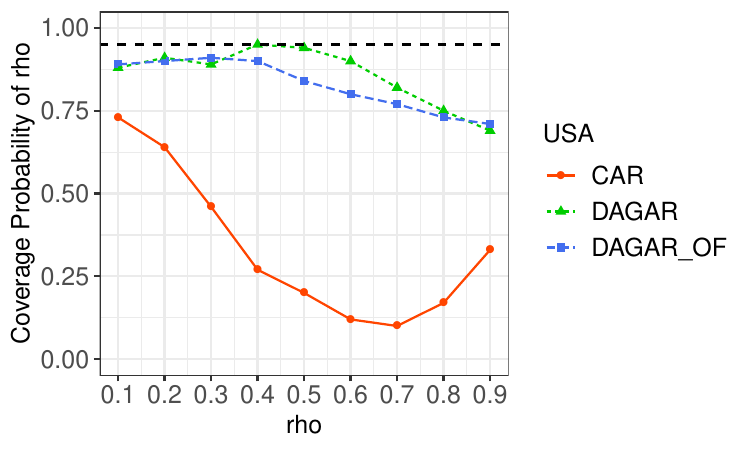}
		\caption{USA: $\rho$}\label{fig:cprhousa}
	\end{subfigure}
	\caption{Coverage probabilities of the parameters as a function of the true $\rho$ (x-axis) for the simulation data analysis using Gaussian data generated from a Mat\'ern$_{3/2}$ GP}\label{fig:mgpcp}
\end{figure}

We first look at the mean square error in terms of estimating the latent spatial random effects in Figure \ref{fig:mgpmse}. We see similar trends as in the case of exponential GP. The MSEs from the ICAR models are much higher, with the scaled ICAR, once again, producing lower MSE than the original ICAR and sparse GLMM. The sparse GLMM was better than the ICAR for path and grid graph but was worse for the USA graph. The proper CAR and the two DAGAR models produced lower MSEs than these ICAR-based models for all three graphs, with the improvement more prominent for smaller $\rho$. For smaller $\rho$, we also see that the DAGAR models produce lowest MSE among all the six models, whereas for larger $\rho$, the MSEs for most of the models are similar. 

We also briefly summarize the comparison of the models based on inference (CP) on the parameters involved. We only look at the common parameters $\beta_1$, $\beta_2$ and $\sigma^2_e$. We do not consider $\rho$ as, unlike the exponential GP,  the spatial decay parameter in the Mat\'ern$_{3/2}$ GP does not have a simple relationship with $\rho$. Figure \ref{fig:mgpcp} provides the coverage probabilities of the three parameters as a function of $\rho$. 
We see once again that the trends observed for the exponential GP data analysis in Section \ref{sec:illu} carry over to here. The coverages for the regression coefficients are close to $95\%$ for all the models except the sparse GLMM. For $\sigma^2_e$, all models produce under-coverage for larger values of $\rho$. For smaller values of $\rho$, however, the coverage of the proper CAR and the two DAGAR models are close to $95\%$. 

\subsection{A non-Gaussian example}\label{sec:ng}  In this Section, we conduct a simulation study using a non-Gaussian response. We generate independent $y_i \sim Poisson(\exp(x_i^\top\beta+w_i))$ where the spatial random effect vector $w=(w_1,w_2,\ldots,w_k)^\top$ are generated as realizations from an exponential GP, akin to Section \ref{sec:illu}. All other parameter and covariate choices remain unchanged from the previous simulations, and the same set of six candidate models are assessed. 

\begin{figure}
	\hskip -2.cm\begin{subfigure}[t]{0.4\textwidth}
		\centering
		\includegraphics[scale=0.5,trim={0.6cm 0cm 3.5cm 0cm},clip]{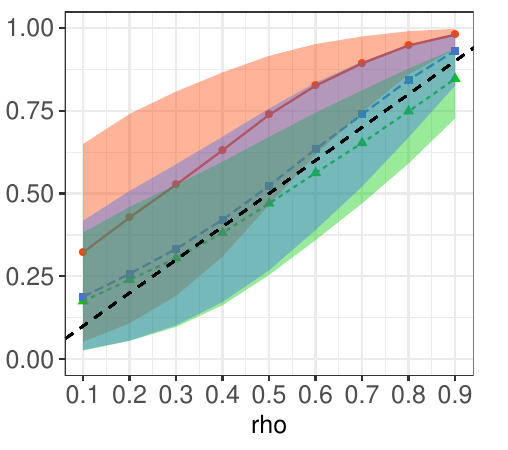}
		\caption{Path}\label{fig:rhoest}
	\end{subfigure} 
	\hskip -2.8cm \begin{subfigure}[t]{0.4\textwidth}
		\centering
		\includegraphics[scale=0.5,trim={1.5cm 0cm 3.5cm 0cm},clip]{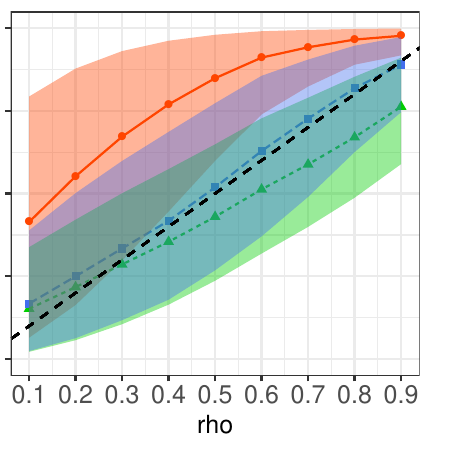}
		\caption{Grid}\label{fig:rhoest}
	\end{subfigure} 
	\hskip -2.3cm \begin{subfigure}[t]{0.4\textwidth}
		\centering
		\includegraphics[scale=0.5,trim={1.5cm 0cm 0cm 0cm},clip]{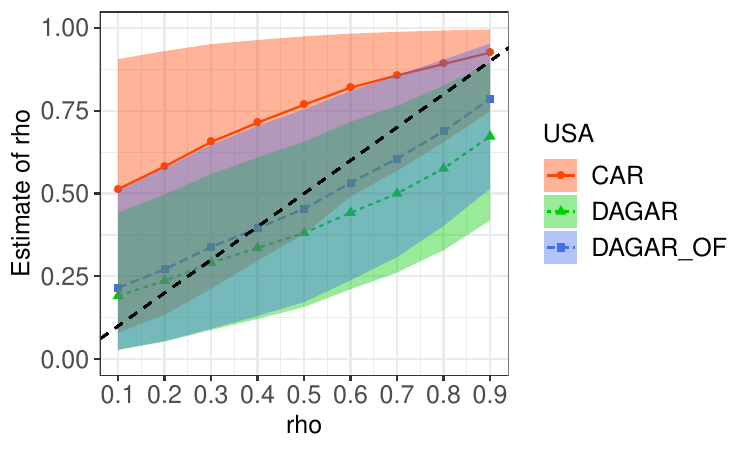}
		\caption{USA}\label{fig:rhoest}
	\end{subfigure}
	\caption{Estimate and confidence bands of $\rho$ as a function of the true $\rho$ (x-axis) for the simulation data analysis using Poisson responses}\label{fig:ngrhoest}
\end{figure}

\begin{figure}[h]
	\centering
	\hskip -2cm\begin{subfigure}[t]{0.4\textwidth}
		\centering
		\includegraphics[scale=0.5,trim={0.6cm 0cm 4.5cm 0cm},clip]{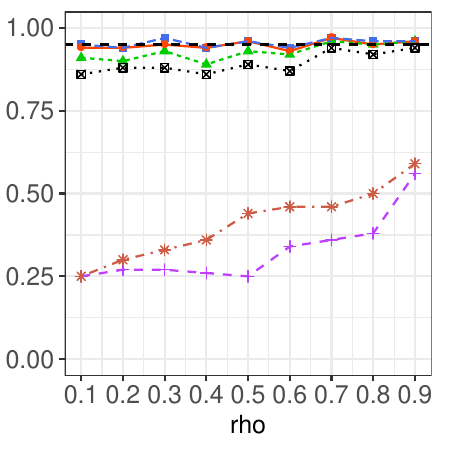}
		\caption{Path: $\beta_1$}\label{fig:cpb1path}
	\end{subfigure}
	\hskip -3cm \begin{subfigure}[t]{0.4\textwidth}
		\centering
		\includegraphics[scale=0.5,trim={0.6cm 0cm 4.5cm 0cm},clip]{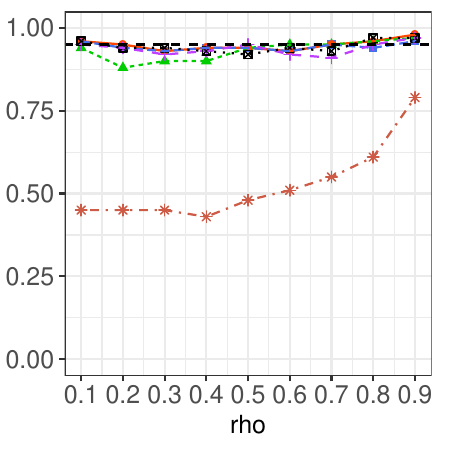}
		\caption{Grid: $\beta_1$}\label{fig:cpb1grid}
	\end{subfigure}
	\hskip -2cm \begin{subfigure}[t]{0.4\textwidth}
		\centering
		\includegraphics[scale=0.5,trim={0.6cm 0cm 0cm 0cm},clip]{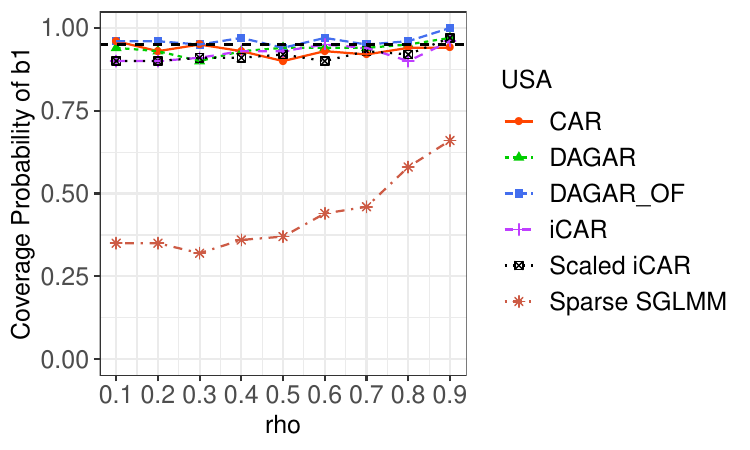}
		\caption{USA: $\beta_1$}\label{fig:cpb1usa}
	\end{subfigure}\\
	\hskip -2cm\begin{subfigure}[t]{0.4\textwidth}
		\centering
		\includegraphics[scale=0.5,trim={0.6cm 0cm 4.5cm 0cm},clip]{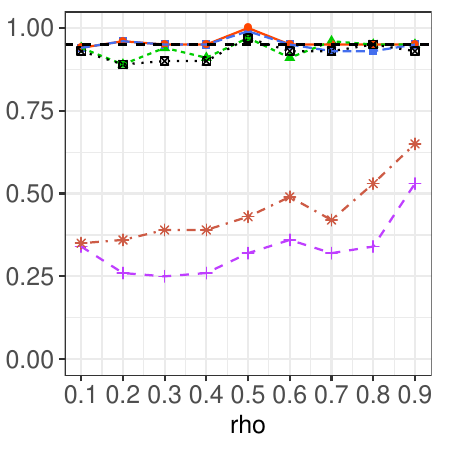}
		\caption{Path: $\beta_2$}\label{fig:cpb2path}
	\end{subfigure}
	\hskip -3cm \begin{subfigure}[t]{0.4\textwidth}
		\centering
		\includegraphics[scale=0.5,trim={0.6cm 0cm 4.5cm 0cm},clip]{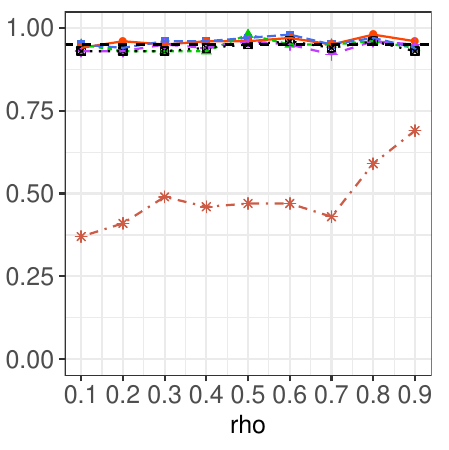}
		\caption{Grid: $\beta_2$}\label{fig:cpb2grid}
	\end{subfigure}
	\hskip -2cm \begin{subfigure}[t]{0.4\textwidth}
		\centering
		\includegraphics[scale=0.5,trim={0.6cm 0cm 0cm 0cm},clip]{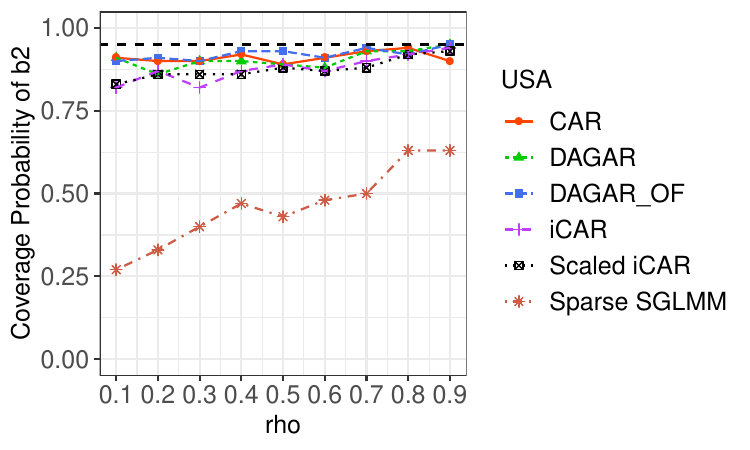}
		\caption{USA: $\beta_2$}\label{fig:cpb2usa}
	\end{subfigure}\\
	\hskip -2cm\begin{subfigure}[t]{0.4\textwidth}
		\centering
		\includegraphics[scale=0.45,trim={0.7cm 0cm 4cm 0cm},clip]{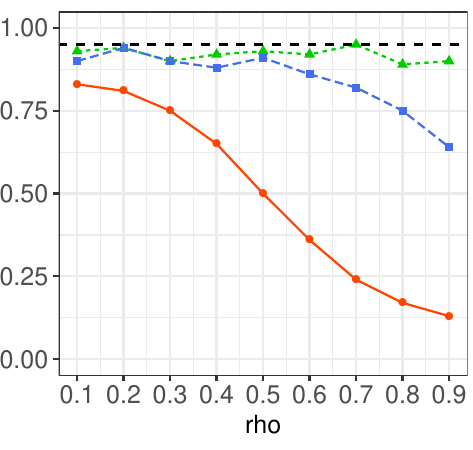}
		\caption{Path: $\rho$}\label{fig:cprhopath}
	\end{subfigure}
	\hskip -3cm \begin{subfigure}[t]{0.4\textwidth}
		\centering
		\includegraphics[scale=0.45,trim={0.7cm 0cm 4cm 0cm},clip]{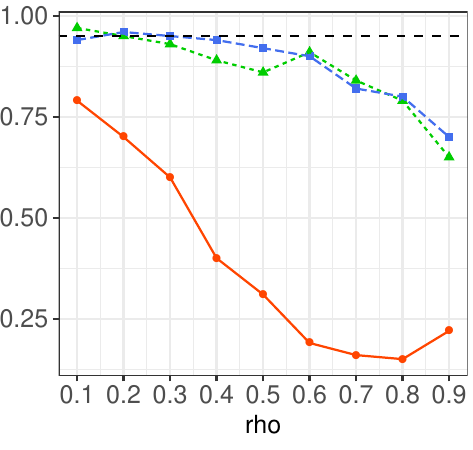}
		\caption{Grid: $\rho$}\label{fig:cprhogrid}
	\end{subfigure}
	\hskip -2cm \begin{subfigure}[t]{0.4\textwidth}
		\centering
		\includegraphics[scale=0.45,trim={0.7cm 0cm 0cm 0cm},clip]{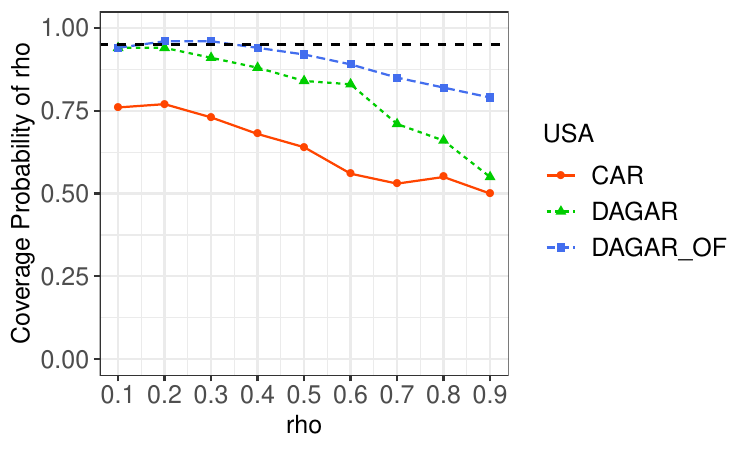}
		\caption{USA: $\rho$}\label{fig:cprhousa}
	\end{subfigure}
	\caption{Coverage probabilities of the parameters as a function of the true $\rho$ (x-axis) for the simulation data analysis using Poisson responses}\label{fig:ngcp}
\end{figure}

We first compare the MSEs which are quite close for all the models except for the sparse GLMM and ICAR (for path graph) which produce significantly higher MSEs. The DAGAR models produced the lowest MSEs for USA graph, and joint lowest MSEs along-with the proper CAR model for the path graph. We then compare the estimation of $\rho$ for the DAGAR and proper CAR models, as for an exponential GP, $\rho$ corresponds to the correlation at unit distance and the data generation ensured that on average the neighboring units are separated by unit distance (see Section \ref{sec:illu}). The estimates and confidence bands in Figure \ref{fig:ngrhoest} demonstrates how the DAGAR model produces accurately estimates the spatial correlation between neighbors even when the data is non-Gaussian, whereas the estimates from the CAR model are far off akin to the Gaussian case. Similarly. the coverage probabilities of parameters in Figure \ref{fig:ngcp}, repeat the trends observed in Figure \ref{fig:gpcp} for the Gaussian case, with all models except the sparse GLMM offering close to $95\%$ coverage for the regression coefficients, and the DAGAR models offering substantially improved coverage for $\rho$ than the proper CAR model. 

\begin{figure}[h!]
	\hskip -2cm\begin{subfigure}[t]{0.4\textwidth}
		\centering
		\includegraphics[scale=0.5,trim={0.6cm 0cm 4.5cm 0cm},clip]{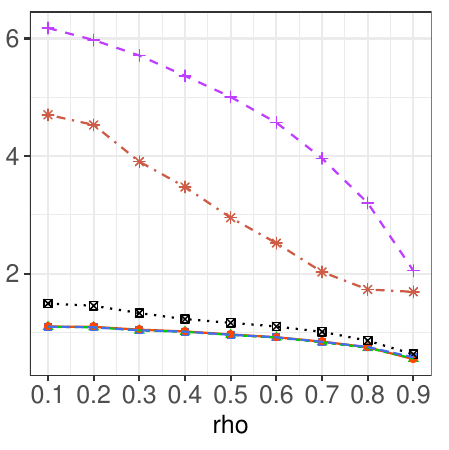}
		\caption{Path}\label{fig:msepath}
	\end{subfigure}
	\hskip -3cm \begin{subfigure}[t]{0.4\textwidth}
		\centering
		\includegraphics[scale=0.5,trim={0.6cm 0cm 4.3cm 0cm},clip]{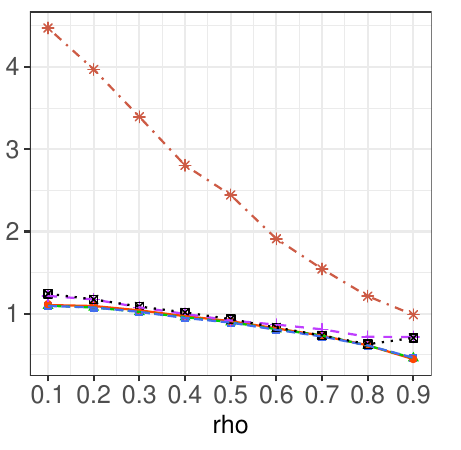}
		\caption{Grid}\label{fig:msegrid}
	\end{subfigure}
	\hskip -2cm \begin{subfigure}[t]{0.4\textwidth}
		\centering
		\includegraphics[scale=0.5,trim={0.6cm 0cm 0cm 0cm},clip]{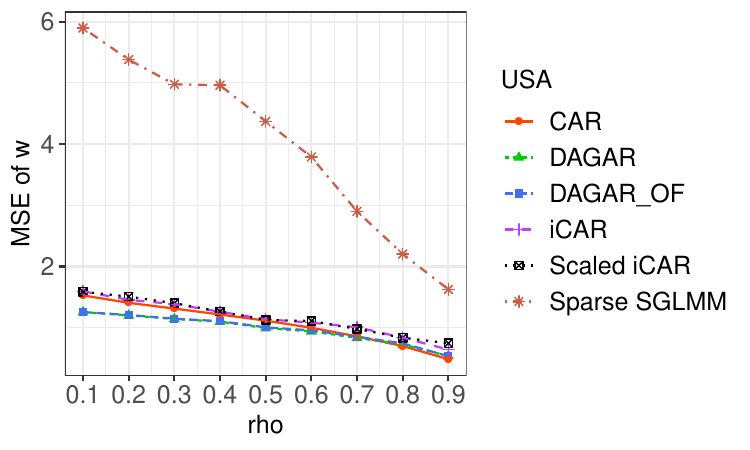}
		\caption{USA}\label{fig:mseusa}
	\end{subfigure}
	\caption{MSE as a function of the true $\rho$ (x-axis) for the simulation data analysis using Poisson responses}\label{fig:ngmse}
\end{figure}

\clearpage

\clearpage
\section{Proofs}\label{sec:proofs}

\subsection{Proof of Theorem \ref{th:intrinsic}}

Let $r=rank(Q)$ and $Q^+$ denote the Moore-Penrose inverse of $Q$. Then, by Theorem 1 part (b) of \citet{higham90}, there exists a permutation $P$ such that 
\[ PQ^+P^\top = R^\top R \mbox{ where } R = \left[ \begin{array}{cc} R_1 & R_2 \\ 0 & 0 \end{array} \right] \]
Here $R_1$ is $r \times r$ upper triangular matrix with positive diagonal elements. Let $D_1 = diag(R_1)$ and $R_1^* = D_1^{-1} R_1$, which has ones on the diagonal.  
We can now write 
\[ R= D U \mbox{ where } U = \left[ \begin{array}{cc} R^*_1 & D_1^{-1} R_2 \\ 0 & I \end{array} \right] \mbox{ and } D = \left[ \begin{array}{cc} D_1 & 0 \\ 0 & 0 \end{array} \right]  \]
Since $U^\top$ is a lower triangular matrix with one on the diagonals, so is $L=U^{-\top}$. Hence, $PQP^\top = (I-B)^\top F (I-B)$ where $F= D^{+2}$ and $B=I-L$ is a strictly lower triangular matrix. 

\subsection{Proof of Theorem \ref{th:pathcov}}

First of all, as $\calT$ is a tree, it is always possible to have an ordering $\pi $ such that $n_\pi (i) = 1$ for any $i \neq \pi (1)$. For example, the orderings corresponding to any pre-order or breadth-first tree traversal of $\calT$ will satisfy this. Without loss of generality we rename the nodes such that $\pi =\{1,\ldots,k\}$ and for $i > 1$, $p(i)$ denotes the directed neighbor of $i$ in $\pi $ implying $p(i) < i$. Letting $w_0=0$, $p(1)=0$ and $n_\pi (1)=0$, the model in ($\ref{eq: nar}$) reduces to $w_i = \rho\;w_{p(i)} +  (1-\rho^2)^{0.5}\eps_i$ where $\eps_i$ are independent standard normal variables. We shall show that for any positive integers $j \leq i \leq k$, $cov(w_i,w_j)=\rho^{d_{ij}}$. We prove this using the strong form of mathematical induction. Since $p(2)=1$, it is easy to verify this for $i=2$. We assume that this is true for $i=2,\ldots,i-1$. It immediately follows that $var(w_i)=\rho^2var(w_{p(i)})+(1-\rho^2)var(\eps_i)=1$. For any $j<i$, $cov(w_i,w_j)=\rho\;cov(w_{p(i)},w_j) = \rho^{1+d_{p(i)j}}$ (by induction). 
Since $\calT$ is acyclic and $n_\pi (i)=1$ for all $i > 1$, the shortest path from $j$ to $i$ runs through $p(i)$. Hence, $d_{ij}=d_{p(i)j}+1$ and the result follows.

\subsection{Proof of Theorem \ref{th:gridcov}}

If $i+j=i'+j'$, then $(i,j)$ and $(i',j')$ are never neighbors. Hence, without loss of generality, we prove the result for $\pi =(S_2,\ldots,S_{m+n})^T$ where $S_r=\{(i,j) \given i+j = r\}$. Let $d((i,j),(i',j'))=|i-i'|+|j-j'|$ denote the Manhattan distance on $\calG$, and  $w_{S_r}$ denote the sub-vector of $w$ corresponding to the indices in $S_r$. It is enough to show by induction on $r$ that $w_{S_r} \sim N(0,(\rho^{D_r})^{-1})$ where $D_{r}$ denotes the distance matrix on $S_r$. This holds trivially for  $r=2$. Let us assume that it holds true for $r-1$. If $(1,r-1) \in S_r$, we define $w(0,r-1) = \rho w(1,r-1) + \eps(0,r-1)$ where $\eps(0,r-1) \sim N(0,1/(1-\rho^2))$ is independent of $w$. If $(r-1,1) \in S_r$ we define $w(r-1,0)$  similarly. Let $w^*_{S_{r-1}}$ be the augmented vector which includes $w(0,r-1)$ or $w(r-1,0)$ or both, along with $w_{S_{r-1}}$. From the construction, $w(0,r-1) = \rho^2 w(1,r-2) + \rho \eps(1,r-1) + \eps(0,r-1)$ implying $var(w(0,r-1))=1$ and $cov(w(0,r-1),w(1,r-2)) =\rho^2$. Hence $cov(w^*_{S_{r-1}})=\rho^{D^*_r}$ where $D^*_r$ is the augmented distance matrix corresponding to $S^*_{r-1}$. Letting $\rho^2=u$, we have for any $(i,j)$ and $(i',j')$ in $S_r$, 
\begin{align*}
cov(w(i,j),&w(i',j'))=\frac{u}{(1+u)^2}\left(cov(w(i-1,j),w(i'-1,j'))+ cov(w(i,j-1),w(i'-1,j')) \right.\\
& + \left.cov(w(i-1,j),w(i',j'-1))+cov(w(i,j-1),w(i',j'-1))\right)+I(i=i')\frac{1-u}{1+u}\\
&=\frac{u}{(1+u)^2}(\rho^{|i-i'+1|+|j-j'-1|}+2\rho^{|i-i'|+|j-j'|}+\rho^{|i-i'-1|+|j-j'+1|})+I(i=i')\frac{1-u}{1+u}
\end{align*}	
If $i=i'$ then $j=j'$ and the expression above equals 1. If $i < i'$, then $j>j'$ and $|i-i'+1|+|j-j'-1|=(i'-i-1)+(j-j'-1)=|i-i'|+|j-j'|-2$. Similarly, $|i-i'-1|+|j-j'+1=|i-i'|+|j-j'|+2$. So, $\rho^{|i-i'+1|+|j-j'-1|}+2\rho^{|i-i'|+|j-j'|}+\rho^{|i-i'-1|+|j-j'+1|} = \rho^{|i-i'|+|j-j'|}(1/u + 2 + u)$. Hence, the results follows.

\subsection{Proof of Theorem \ref{th:main}}\label{app:proof}
For any vertex $i$ with $n_i$ neighbors, let $\pi _{ir}$ denote the set of all permutations $\pi $ such that $n_{\pi (i)}=r$. By symmetry, $|\pi _{ir}|=k!/(n_i+1)$ for $r=0,1,\ldots,n_i$. Also, for any $i \sim j$ and $r=0,\ldots,n_i$, let $\pi _{ijr}$ denote the set of all permutations $\pi $ such that $n_{\pi (i)}=r$ and $j \in N_\pi (i)$. Then,  $|\pi _{ijr}|= k!/(n_i+1)\times pr(j$ is among the $r$ directed neighbors of $i) =  rk!/(n_i(n_i+1))$. We now have
\begin{align*}
Q[i,i] &= \frac 1{k!(1-\rho^2)} \sum_\pi  \left( 1+(n_{\pi (i)}-1)\rho^2 + \sum_{j \sim i} I(i \in N_\pi (j)) \frac {\rho^2}{1+(n_{\pi (j)}-1)\rho^2} \right) \\
&= 1 + \frac {\rho^2}{k!(1-\rho^2)} \left( \sum_{r=0}^{n_i}  r |\pi _r| + \sum_{j \sim i} \sum_{r=0}^{n_j}  \frac {|\pi _{jir}|}{1+(r-1)\rho^2} \right) \\
&= 1 + \frac {n_i\rho^2}{2(1-\rho^2)} + \frac {\rho^2}{1-\rho^2} \sum_{j \sim i} \frac 1{n_j(n_j+1)} \sum_{r=1}^{n_j} \frac{r}{1+(r-1)\rho^2}\\
&= 1 + \frac {n_i\rho^2}{2(1-\rho^2)} + \frac {\rho^2}{1-\rho^2} \sum_{j \sim i} \frac 1{n_j(n_j+1)} f(\rho,n_j).
\end{align*}
To evaluate the non-diagonal entries of $Q$, we additionally define $\pi _{ijkr}$ to be the set of all permutations $\pi $ such that $n_{\pi (i)}=r$ and $\{j,k\} \subseteq N_\pi (i)$. Applying the combinatorial argument used earlier, we see that $|\pi _{ijkr}|=r(r-1)k!/((n_i-1)n_i(n_i+1))$. Let $i \approx j$ implies that there exists at least one node $k$ such that $i \sim k$ and $j \sim k$. 
\begin{align*}
Q[i,j] &= \frac 1{k!(1-\rho^2)} \sum_\pi  \left( -\rho I(i \sim j) + I(i \approx j) \sum_{k: \{i,j\} \subseteq N_\pi (k)} \frac {\rho^2}{1+(n_{\pi (k)}-1)\rho^2} \right) \\
&= -\frac{\rho}{1-\rho^2}I(i \sim j) + \frac {\rho^2}{k!(1-\rho^2)} I(i \approx j) \sum_{k \sim N(i) \cap N(j)} \sum_{r=0}^{n_k}  \frac {|\pi _{kijr}|}{1+(r-1)\rho^2}  \\
&= -\frac{\rho}{1-\rho^2}I(i \sim j) + \frac {\rho^2}{1-\rho^2}I(i \approx j) \sum_{k \sim N(i) \cap N(j)} \frac 1{(n_k-1)n_k(n_k+1)}\sum_{r=1}^{n_k} \frac{r(r-1)}{1+(r-1)\rho^2}\\
&= -\frac{\rho}{1-\rho^2}I(i \sim j) + \frac {1}{1-\rho^2}I(i \approx j) \sum_{k \sim N(i) \cap N(j)} \left( \frac 1{2(n_k-1)} - \frac 1{(n_k-1)n_k(n_k+1)}f(\rho,n_k) \right).
\end{align*}

\subsection{Proof of Theorem \ref{th:frobpath}}
We write $Q_\pi (\rho)$ and $Q(\rho)$ as $Q_\pi $ and $Q$ hiding the dependence on $\rho$ except when necessary. From Theorem \ref{th:main} we have for $i=3,\ldots,k-2$, $Q_{ii} = \frac{3 + 6\rho^2  + \rho^4}{3(1+\rho^2)(1-\rho^2)}$, $Q_{i,i+1} = -\frac \rho{1-\rho^2}$ and $Q_{i,i+2} = \frac {\rho^2} {3(1+\rho^2)(1-\rho^2)}$. Hence, 
\[ ||Q ||_F^2  = \frac k {9(1+\rho^2)^2(1-\rho^2)^2} \left((3 + 6\rho^2 + \rho^4)^2 + 18\rho^2(1+\rho^2)^2 + 2\rho^4 \right) + o(k) \]
where the $o(k)$ term arises from rows and columns corresponding to the nodes at the extreme right or left. 

Now using left-to-right or right-to-left ordering, from (\ref{eq:btau}),  a typical term in the quadratic form $w'Q_\pi  w$ will be of the form $(w_i - \rho w_{i-1})^2/(1-\rho^2)$. Hence, for $i=1,2,\ldots,k-2$, $Q_{\pi:ii} = \frac {1+\rho^2}{1-\rho^2}$, $Q_{\pi:i,i+1} = -\frac \rho {1-\rho^2}$ and $Q_{\pi:i,i+2} = 0$. So,
\begin{align*}
||Q - Q_\pi  ||_F^2  & = \frac k {9(1+\rho^2)^2(1-\rho^2)^2} \left((3 + 6\rho^2 + \rho^4 - 3(1+\rho^2)^2)^2 + 2\rho^4 \right) + o(k) \\
& = \frac k {9(1+\rho^2)^2(1-\rho^2)^2} (4\rho^8 + 2\rho^4) + o(k)
\end{align*}
Hence, the result follows. 

\subsection{Proof of Theorem \ref{th:frobgrid}}
We index the nodes of the grid as $(i,j)$, and entries of $Q_\pi $ and $Q$ as $Q_{\pi :(ij),(i'j')}$ and $Q_{(ij),(i'j')}$ respectively for $1 \leq  i,i',j,j' \leq  m$. Like in the proof of Theorem \ref{th:frobpath}, it only suffices to evaluate the Frobenius norms for the interior points of the grid (having $4$ neighbors each of whom also have $4$ neighbors) as the contribution from the remaining terms will be $o(m^2)$. Hence from Theorem \ref{th:gridcov} we have for $3 \leq  i,i',j,j' \leq  m-2$, 
\begin{align*}
Q_{(ij),(ij)} & = \frac {1+ \rho^2 + \rho^2 s(\rho)/5}{1-\rho^2} \\
Q_{(ij),(i+1,j)} & = Q_{(ij),(i,j+1)} = - \frac \rho {1- \rho^2} \\
Q_{(ij),(i+2,j)} & = Q_{(ij),(i,j+2)} = \frac 1 {1-\rho^2} (1/6 - s(\rho)/60)\\
Q_{(ij),(i+1,j+1)} & = Q_{(ij),(i+1,j-1)} = \frac 2 {1-\rho^2} (1/6 - s(\rho)/60)
\end{align*}
Summing up, we have
\begin{align*}
||Q||_F^2 & = m^2 (Q_{(ij),(ij)}^2 + 4(Q_{(ij),(i+1,j)}^2 + Q_{(ij),(i+2,j)}^2 + Q_{(ij),(i+1,j+1)}^2) +o(1))\\
& = \frac {m^2}{(1-\rho^2)^2} \left((1+ \rho^2 + \rho^2 s(\rho)/5)^2 + 4\rho^2 + 20 (1/6 - s(\rho)/60)^2 \right) + o(m^2)
\end{align*}

Now, without loss of generality we assume that the DAGAR precision matrix $Q_\pi (\rho)$ was constructed by ordering the nodes in increasing order of $(i+j)$. Then, a typical term in the quadratic form $w'Q_\pi  w$ will be of the form $\frac {1+ \rho^2}{1-\rho^2}(w_{ij} - \frac \rho{1+\rho^2}(w_{i,j-1}+w_{i-1,j}))^2$. 
Hence, we will have 
\begin{align*}
Q_{\pi :(ij),(ij)} & = \frac {1+ \rho^2 + 2\rho^2/(1+\rho^2)}{1-\rho^2} \\
Q_{\pi :(ij),(i+1,j)} & = Q_{\pi :(ij),(i,j+1)} = - \frac \rho {1- \rho^2} \\
Q_{\pi :(ij),(i+2,j)} & = Q_{\pi :(ij),(i,j+2)} = Q_{\pi :(ij),(i+1,j+1)} = 0\\
Q_{\pi :(ij),(i+1,j-1)} & = \frac {\rho^2} {(1+\rho^2)(1-\rho^2)}
\end{align*}
Subtracting, we have
\begin{align*}
||Q_\pi  - Q||_F^2 = m^2 & ((Q_{\pi :(ij),(ij)}-Q_{(ij),(ij)})^2 + 4 Q_{(ij),(i+2,j)}^2 + \\
&  2 (Q_{\pi :(ij),(i+1,j-1)}-Q_{(ij),(i+1,j-1)})^2 + \\
& 2 Q_{(ij),(i+1,j+1)}^2 +o(1))
\end{align*}
Hence, the result follows.

\bibliographystyle{apalike}
\bibliography{references}

\label{lastpage}
\end{document}